\documentstyle[aps,amssymb,amsfonts,epsf,multicol,prb]{revtex}
\renewcommand{\narrowtext}{\noindent\begin{multicols}{2}\noindent
\global\columnwidth20.5pc}
\renewcommand{\widetext}{\end{multicols}
\global\columnwidth42.5pc}  
\renewcommand{\top}[1]{%
 \vskip #1%
 \begin{picture}(290,80)(80,500)%
 \thinlines%
 \put(65,500){\line( 1, 0){255}}\put(320,500){\line( 0, 1){5}}%
 \end{picture}%
}
\newcommand{\bottom}[1]{%
 \vskip #1%
 \begin{picture}(290,80)(80,500)%
 \thinlines%
 \put(330,500){\line( 1, 0){255}}\put(330,500){\line( 0, -1){5}}%
 \end{picture}%
}
\input{psfig.sty}

\begin{document}
\def\lsim{\buildrel <\over\sim}
\def\gsim{\buildrel >\over\sim}
{
\title{Fermi and Non-Fermi Liquid Behavior of Quantum Impurity
Models:\footnote{Dedicated to Erwin M\"uller-Hartmann on the
occasion of his 60th birthday}\\
A Diagrammatic Pseudo-particle Approach}

\author{J. Kroha and P. W\"olfle}

\address{Institut f\"ur Theorie der Kondensierten Materie,
Universit\"at Karlsruhe, 76128 Karlsruhe, Germany}

\maketitle

\begin{abstract}
We review a systematic many-body method capable of describing Fermi
liquid and Non-Fermi liquid behavior of quantum impurity models at low
temperatures on the same footing.  The crossover to the high
temperature local moment regime is covered as well.  The approach
makes use of a pseudo-particle representation of the impurity Hilbert
space in the limit of infinite Coulomb repulsion $U$ as well as for
finite $U$.  Approximations are derived from a generating
Luttinger-Ward functional, in terms of renormalized perturbation theory
in the hybridization $V$.  Within a ``conserving T-matrix
approximation''
(CTMA),  including all two-particle vertex functions, an infinite
series of leading infrared singular skeleton diagrams is included.
The local constraint is strictly enforced. Applied to the SU(N) $\times$
SU(M) multichannel Anderson model the method allows to recover the
Fermi liquid behavior of the single channel case, as well as the
non-Fermi liquid behavior in the multi-channel case.  The results are
compared with the ``non-crossing approximation'' (NCA) and with data
obtained by the numerical renormalization group and the Bethe ansatz.  
The generalization of the method to the case of finite on-site
repulsion $U$ is presented in a systematical way and solved on the
level of a generalized NCA, fully symmetric with respect to all virtual 
excitations of the model.
\end{abstract}
}
\narrowtext

\section{Introduction}

Over the past two decades the problem of correlated electrons on a
lattice has emerged as a central theme of condensed matter theory.
Since the mid seventies numerous metallic systems have been found in
which the conduction electron system is believed to be dominated by a
strong on-site Coulomb interaction.  Examples are the heavy fermion
compounds \cite{Lee86} and the cuprate superconductors \cite{Ander97},
but systems like the two-dimensional electron gas in quantum Hall
devices \cite{Sarma96} are governed by similar physics.  
Typical model Hamiltonians are the
Hubbard model and the t-J model for the single-band situation, and the
Anderson or Kondo lattice models in the multi-band case.  In the event
that the local Coulomb repulsion in these models exceeds the band
width, conventional many-body theory, i.e. perturbation theory
(including infinite resummations) in the Coulomb repulsion does not
work anymore.  The obvious alternative, perturbation expansion in
terms of the kinetic energy meets with severe difficulty, mainly due
to the infinite degeneracy of the ground state of a lattice model in
the limit of zero hopping (a collection of independent atoms) and 
the non-canonical commutation relations of the field operators 
of the eigenstates in the atomic limit. It is
therefore desirable to develop new methods which are capable of
dealing with these problems.  

To make it clear what is the principal obstacle for applying the
well-established many-body techniques to this class of problems let us
consider the Hubbard model with large on-site repulsion $U$, where
each lattice site can either be empty (state $\mid 0 \rangle$), singly occupied
($\mid\uparrow \rangle, \mid\downarrow \rangle$ for an electron with spin
projection $\uparrow, \downarrow$) or doubly occupied $(\mid 2 \rangle)$.
The dynamics of an electron will be very different according to
whether it resides on a singly or doubly occupied site.  For less than
half band filling and large
repulsive $U$ the doubly occupied states will be very costly in
energy, and will contribute to the low energy physics only via virtual
processes. This is the typical feature of strongly correlated
electrons: the dynamics is constrained to a subspace of the total
Hilbert space.  The necessary projection onto this subspace is
difficult to perform within conventional many-body theory and requires
the development of new theoretical tools. This article deals with the
method of auxiliary particles introduced to effect the projection in
Hilbert space, while keeping most of the desirable features of
renormalized perturbation theory.  

The idea of auxiliary particles is simple: for each of the four states
$\mid 0 \rangle, \mid \uparrow \rangle, \mid \downarrow \rangle, \mid
2\rangle$  at a given
lattice site (considering one orbital per site), a particle is
introduced, which is created out of a vacuum state without any lattice
site at all.  The Fermi character of the electron requires that two of
the auxiliary particles are fermions, e.g. the ones representing
$\mid\uparrow \rangle$ and $\mid\downarrow \rangle$, and the remaining two are
bosons.  To ensure that a lattice site is in one of the four
physical states, the number of auxiliary particles is constrained to be
exactly one.  Compared to alternative ways of effecting the projection
in Hilbert space, the auxiliary particle method has the advantage of
allowing one to use the machinery of quantum field theory, provided
the constraint can be dealt with in a satisfactory way. As will be
shown below, the constraint is closely related to a fictitious gauge
symmetry in the Fock space of pseudo particles.  The formulation in
terms of renormalized perturbation theory in the kinetic energy
presented in this paper allows to conserve the gauge symmetry in any
chosen conserving approximation.  Conserving approximations, on the
other hand, are generated by functional derivation of a Luttinger-Ward
functional.  The Luttinger-Ward functional is chosen taking into
account information on important physical processes  as well as using
expansion in a small parameter, e.g. the hybridization energy in units
of the conduction electron band width. Auxiliary particle
representations were pioneered by Abrikosov \cite{Abrik65}, who first
introduced a fermionic representation for local spins, and later by
Barnes \cite{Barnes76}, who was the first to define and use the pseudo
particle representation of a local impurity level we will be working
with below.

Somewhat later these representations were applied to the N-fold
degenerate Anderson-impurity models in conjunction with a mean-field
approximation (MFA) of the slave boson fields \cite{Coleman84,Newns83}.
As will be discussed below, the MFA, inspite of its initial success,
suffers from severe problems: (1) it leads invariably to the Fermi
liquid ground state, even for multi-channel models, in contradiction
to the exact non-Fermi liquid behavior in those cases; (2) even for the
single--channel case the MFA features a spurious first order transition
separating the low temperature Kondo screened state from the high
temperature local moment regime.  On the technical side, the MFA
violates the gauge symmetry mentioned above.  The latter problem 
may be corrected by
including the leading fluctuations around the saddle-point in $1/N$
expansion \cite{Newns83}, whereas the former problems remain. We shall
therefore avoid using the mean-field approximation which solves the
problem of the spurious phase transition.

\section{Single- and multi-channel quantum impurity models}
\label{single}

\begin{figure}
\vspace*{-0cm}
\centerline{\psfig{figure=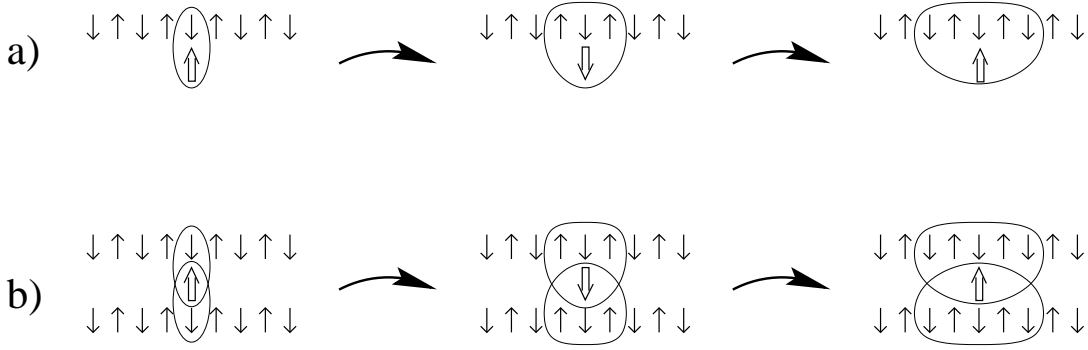,width=0.8\textwidth}}
\vspace*{0.6cm}
\caption{
Sketch of the renormalization group for a) the single--channel Kondo 
model
(local moment compensation) and 
b) the two--channel Kondo model (local moment over--compensation). 
Small arrows  denote conduction electron spins 1/2, a heavy
arrow a localized spin 1/2. The curved arrows indicate successive 
renormalization steps.
\label{cartoon}}
\end{figure}

Quantum impurity models such as the Kondo and Anderson impurity models
were discussed first in the context of magnetic impurities in metals
\cite{Hewson93}. In these models an N-fold spin degenerate local
impurity level (here called d-level) hybridizes with the conduction
band.  In the presence of a sufficiently strong Coulomb repulsion $U$
between any two electrons in the local impurity level and provided the
level lies sufficiently far below the Fermi energy the local level will
be singly occupied and a magnetic moment forms. The antiferromagnetic
exchange interaction between the local moment and the local conduction
electron spin density leads to the Kondo effect.  The physics of the
Kondo effect is not restricted to magnetic ions in metals.  It depends
only on the local level being degenerate (any two-level or multi-level
system qualifies) and the local Coulomb interaction $U$ being
sufficiently strong.  The Kondo effect can therefore occur as well for
quadrupole moments, the two-level systems in amorphous metals 
\cite{Cox98} or for
the energy level system of quantum dots \cite{schoen97}. We will
consider the prototype Anderson impurity Hamiltonian \cite{Ander61}
\begin{equation}
H = H_c + H_d + H_{hyb}
\label{2.1}
\end{equation}
with the conduction electron Hamiltonian
\begin{equation}
H_c = \sum_{\vec k,\sigma} \epsilon_k c_{\vec k\sigma}c_{\vec
k\sigma}\;\;\;;\;\;\;\;\sigma = 1,\ldots N
\label{2.2}
\end{equation}
the local d-electron Hamiltonian
\begin{equation}
H_d = E_d \sum_{\sigma} d_\sigma^+d_\sigma + U \sum_{\sigma < 
\sigma'}n_{d\sigma}n_{d\sigma{'}}
\label{2.3}
\end{equation}
and the hybridization term
\begin{equation}
H_{hyb} = V \sum_{\vec k\sigma}(c_{\vec k\sigma}^+ d_\sigma + h.c)
\label{2.4}
\end{equation}
In lowest order perturbation theory in $V$ the d-level at energy $E_d$
acquires a width $\Gamma = \pi V^2{\cal N}(0)$, with ${\cal N}(0)$ the 
conduction
electron density of states at the Fermi level $\omega = 0$.  In the
Kondo limit of the model, defined by $E_d < 0$, $2 E_d + U >
0$, $\Gamma/\mid E_d\mid \ll 1$ and $\Gamma /(2E_d+U) \ll 1$, the low
energy sector of the Anderson model may be mapped onto the Kondo model
\cite{Schrief66} featuring a local antiferromagnetic exchange
interaction $J$ of the local spin of the single electron in the
d-level with the local conduction electron spin, where $J=V^2 [1/\mid
E_d\mid + 1/(2E_d+U)]$. 
These models have been studied extensively by means
of Wilson's renormalization group \cite{Wilson75}, by the Bethe 
ansatz method \cite{Andrei83,Wieg83} and by means of a 
phenomenological Fermi liquid theory \cite{Nozieres74}. 
In this way the following physical 
picture has emerged \cite{Hewson93}: 
The model contains a dynamically generated low 
temperature scale, the Kondo temperature, which, for $U\to \infty$,
is expressed in terms 
of the parameters of the Anderson Hamiltonian  as
\begin{equation}
T_K = D (N \Gamma / D )^{(M/N)}{\rm e}^{-\pi E_d / N \Gamma }, 
\label{2.4a}
\end{equation}
with ${\cal N}(0)$ and $D=1/{\cal N}(0)$
the density of states at the 
Fermi energy and the high energy band cutoff, respectively.
$N$, $M$ are the degeneracy
of the local level and the number of conduction electron channels 
(see below). In the intermediate temperature regime, $T \gsim T_K$,
resonant spin flip scattering of electrons at the Fermi surface 
off the local degenerate level leads to logarithmic corrections to the 
magnetic susceptibility $\chi(T)$, the linear specific heat
coefficient $\gamma (T)$  and 
the resistivity $\rho (T)$ 
and to a breakdown of perturbation theory 
at $T\simeq T_K$. In the single--channel case $(M=1)$, below $T_K$ a 
collective many--body spin singlet 
state develops in which the impurity spin is screened by 
the conduction 
electron spins as lower and lower energy scales are successively 
approached, leaving the system with a pure potential scattering center. 
The spin singlet formation is sketched in Fig.~\ref{cartoon} a) 
and corresponds to a vanishing entropy at $T=0$, $S(0)=0$. 
It also leads to saturated behavior of physical quantities 
below $T_K$, like $\chi (T)=const$,$\gamma(T) =  c (T)/T =const.$ 
and $\Delta \rho (T) = \rho (T) - \rho(0) \propto T^2$, i.e. to Fermi liquid 
behavior.

\widetext
\narrowtext

Multichannel quantum impurity models are characterized by the
completely symmetric coupling of $M$ degenerate conduction bands to
the impurity.  Let us assume for definiteness that the ground state of
the impurity is a magnetic singlet, with $M$-fold orbital degeneracy,
$\mid\Gamma\mu >$, where $\mu = 1,\ldots M$, and the first excited
state, obtained by adding one electron to the impurity, 
is magnetic and N-fold degenerate, but is an orbital singlet,
$\mid\Gamma{'}\sigma >$, $\sigma = 1,\ldots, N$.  An SU(N) $\times$
SU(M) symmetric Anderson model may then be defined in the limit
$U\rightarrow \infty$ as \cite{Cox98}
\begin{equation}
H = H_{c\mu}= + E_d \sum_\sigma \mid
\Gamma{'}\sigma\rangle\langle\Gamma{'}\sigma\mid + H_{hyb}
\label{2.5}
\end{equation}
where
\begin{equation}
H_{c\mu}=
\sum_{\vec{k}\sigma} \varepsilon _{\vec{k}}
c_{\vec{k}\mu\sigma}^{\dag}
c_{\vec{k}\mu\sigma}
\label{2.5a}
\end{equation}
describes the conduction electron continuum in orbital channel $\mu$ and
\begin{equation}
H_{hyb} = V \sum_{\vec k,\sigma,\mu} \Big(c_{\vec k\mu\sigma}^+ \mid
\Gamma \mu \rangle\langle \Gamma{'}\sigma\mid + h.c.\Big)
\label{2.6}
\end{equation}
and the effect of the Coulomb interaction is to restrict the average
occupation to $n_d =\sum_\sigma \mid\Gamma{'}\sigma\rangle\langle
\Gamma{'}\sigma\mid \le 1$.

Like in the single channel case, in the multi--channel case, too, 
the conduction electrons of each channel {\it separately} screen the 
impurity moment by multiple spin scattering at temperatures below the 
Kondo scale $T_K$. However, in this case, the local moment is 
over--compensated, since the impurity spin can never form a singlet 
state with both conduction electron channels at the same time in this 
way, as can be seen in Fig.~\ref{cartoon} b). As a consequence of 
this frustration, there is not a unique ground state, leading to a 
finite residual entropy \cite{Andrei84,Tsvelik85} at $T=0$ of 
$S(0)=k_B {\rm ln}\sqrt{2}$ in the two--channel model. In particular, 
the precondition of FL theory of a 1:1 correspondence 
between interacting and non--interacting states is violated. 
As a consequence, characteristic singular temperature dependence 
\cite{Andrei84,affleck.91} of physical quantities persists 
for $T\lsim T_K$ down to $T=0$: $\chi (T)\propto -{\rm ln}(T/T_K)$, 
$c (T)/T  \propto -{\rm ln}(T/T_K)$ and 
$\rho (T)-\rho(0)\propto -\sqrt {T/T_K}$.

\section{Pseudo-particle representation}
\label{pseudo}

As discussed above, the dynamics of an electron occupying a local
level of a quantum impurity will depend in an essential way on whether
the level is singly occupied or multiply occupied, provided the
Coulomb interaction $U$ between two electrons in the same level is
large. It is therefore useful to divide the Hilbert space into sectors
labeled by the occupation number of the quantum impurity. For each
Fock state $\mid\alpha >$ of the impurity one may define a creation
operator $a_\alpha^+$, which, when operating on a vacuum state $\mid
{\rm vac} >$, creates the state $\mid \alpha >\;\;:\;\;\mid\alpha> =
a_\alpha^+\mid {\rm vac}>$.  For a usual impurity level with spin 
degeneracy $N=2$ one thus defines two bosons $b$ and $a$, and two
fermions $f_\sigma$, $\sigma = \uparrow,\downarrow$, creating the
empty and doubly occupied states $\mid 0 > = b^+\mid {\rm vac} >$,
$\mid 2 > = a^+\mid {\rm vac} >$, and the singly occupied states
$\mid\sigma > = f_\sigma^+ \mid {\rm vac}>$. The creation of an
electron in the empty level, effected by a fermion creation operator
$d_\sigma^+$, according to $\mid\sigma > = d_\sigma^+ \mid 0 >$ or
$\mid 2 > = d_\uparrow^+ \mid\downarrow > = - d_\downarrow^+
\mid\uparrow >$, is described in terms of the pseudo-particle operators
as
\begin{equation}
d_\sigma^+ = f_\sigma^+ b +\eta_\sigma a^+f_{-\sigma}
\label{3.1}
\end{equation}
where the factor $\eta_\sigma = \pm 1$ for $\sigma =
\uparrow,\downarrow$ accounts for the antisymmetry of the doubly
occupied state.  By the introduction of the pseudo-particles the Fock
space has been vastly extended.  The physical sector of this
artificial Fock space is defined by the requirement that the impurity
should be occupied by exactly one pseudo-particle at any time
(corresponding to a definite state), as expressed by the constraint
\begin{equation}
\sum_\sigma f_\sigma^+f_\sigma + b^+b + a^+a = 1
\label{3.2}
\end{equation}
In the following we will largely consider the case of infinite Coulomb
repulsion $U$, implying that double (multiple) occupancy of the
impurity is not possible.  The operators $a^+$, $a$ will therefore not
appear in this case.  In the multi-channel SU(N) $\times$ SU(M)
Anderson model discussed in the last section, we will, however, need
to introduce $M$ boson operators $b_\mu^+$, $\mu = 1,\ldots M$ creating
the $M$ ground states $\mid\Gamma\mu >$, and $N$ fermion operators
$f_\sigma^+$, $\sigma = 1,\ldots N$. 
In terms of these
operators the Hamiltonian of the SU(N)$\times$SU(M) Anderson model 
 takes the form
\begin{eqnarray}
H&=&\sum _{\vec k,\sigma ,\mu}\varepsilon _{\vec k}
c_{\vec k\mu\sigma}^{\dag}c_{\vec k\mu\sigma}+
E_d\sum _{\sigma} f_{\sigma}^{\dag}f_{\sigma} \nonumber\\
&+&V\sum _{\vec k,\sigma ,\mu}(c_{\vec 
k\mu\sigma}^{\dag}b_{\bar\mu}^{\dag}f_{\sigma} +h.c.)
\label{sbhamilton}
\end{eqnarray}
In order for $H$ to be SU(M) invariant, the
slave boson multiplet $b_{\bar\mu}$ transforms according to the conjugate 
representation of the SU(M). In addition, the operator constraint
\begin{equation}
Q\equiv \sum _{\sigma} f_{\sigma}^{\dag}f_{\sigma} +
        \sum _{\mu} b_{\bar\mu}^{\dag}b_{\bar\mu} =1
\end{equation}
has to be satisfied at all times. One might interpret the constraint as
a statement of charge quantization, with the integer $Q$ the conserved,
quantized charge. Similar to quantum field theories with conserved 
charges, the charge conservation is intimately related to the existence 
of a local gauge symmetry. Indeed, the system defined by the Hamiltonian 
Eq.~(\ref{sbhamilton}) is invariant under simultaneous local $U(1)$ 
gauge transformations $f_{\sigma}\rightarrow f_{\sigma} {\rm e}^{i\phi 
(\tau )}$, $b_{\bar\mu}\rightarrow b_{\bar\mu} {\rm e}^{i\phi (\tau )}$, with 
$\phi (\tau )$ an arbitrary time dependent phase.

\subsection{Exact projection onto the physical Hilbert space}
While the gauge symmetry guarantees the conservation of the quantized 
charge $Q$, it does not single out any particular $Q$, such as $Q=1$. 
In order to effect the projection onto the sector of Fock space 
with $Q=1$, one may use a procedure first proposed by
Abrikosov \cite{Abrik65}: Consider first the grand--canonical 
ensemble with respect to $Q$, defined by the statistical operator
\begin{equation}
\hat \rho _G = \frac{1}{Z_G} {\rm e}^{-\beta (H+\lambda Q)},
\label{eq:stat_op}
\end{equation}
where $Z_G={\rm tr}[{\rm exp}\{-\beta (H+\lambda Q)\}]$ 
is the grand--canonical partition function with respect to
$Q$, $-\lambda$ is the associated chemical potential,
and the trace extends over the complete Fock space,
including summation over $Q$. The expectation value of an observable
$\hat A$ in the grand--canonical ensemble is given by
\begin{equation}
\langle \hat A\rangle _G = {\rm tr} [\hat \rho _G \hat A] .
\label{eq:exp_val}
\end{equation}
The physical expectation value of $\hat A$, $\langle \hat A\rangle$, 
is to be evaluated in the canonical ensemble where $Q=1$. It can be 
calculated from the grand--canonical ensemble by differentiating with 
respect to the fugacity $\zeta = {\rm e}^{-\beta\lambda}$ and
taking $\lambda$ to infinity \cite{Coleman84},
\begin{equation}
\langle \hat A\rangle =
\lim _{\lambda \rightarrow \infty}
\frac {\frac{\partial }{\partial \zeta} \mbox{tr}
       \bigl[\hat A e^{-\beta (H+\lambda Q)} \bigr]} 
      {\frac{\partial }{\partial \zeta} \mbox{tr}
       \bigl[ e^{-\beta (H+\lambda Q)} \bigr]} =
\lim _{\lambda \rightarrow\infty}\frac{\langle Q\hat A\rangle _G}
{\langle Q \rangle _G}\ .
\label{projection}  
\end{equation}

{\em Projecting operators acting on the impurity states:}\/--- 
We list two important results, which follow straightforwardly
from Eq. (\ref{projection}): First,
the canonical partition function in the subspace $Q=1$ is
\begin{eqnarray}
Z_C & = & \lim _{\lambda \rightarrow \infty} \mbox{tr}
          \bigl[ Q e^{-\beta (H+\lambda (Q-1))} \bigr] \nonumber \\ 
    & = & \lim _{\lambda \rightarrow \infty}
          \bigl( e^{\beta\lambda} \langle Q \rangle _{G}(\lambda ) 
          \bigr)Z_{Q=0}\ \ \ ,\label{eq:zcan}
\end{eqnarray}
where the subscripts $G$ and $C$ denote the grand--canonical and the
canonical ($Q=1$) expectation value, respectively. Second, the 
canonical $Q=1$ expectation
value of any operator $\hat A$ which has a zero expectation value in 
the $Q=0$ subspace, $\hat A|Q=0\rangle =0$, is given by,
\begin{equation}
\langle \hat A \rangle _C = \lim _{\lambda \rightarrow \infty}
\frac {\langle \hat A \rangle _{G}(\lambda )}{ 
\langle Q \rangle _{G}(\lambda )}
\label{eq:canonical-expectation}
\end{equation}
Note that $\hat A|Q=0\rangle =0$ holds true for any physically observable
operator acting on the impurity. Examples are the physical electron operator 
$d^{\dag}_{\mu\sigma}=f^{\dag}_{\sigma}b_{\bar\mu}$ or the local spin
operator 
$\vec S = \sum _{\sigma\sigma '}
\frac{1}{2}f^{\dag}_{\sigma}\vec \tau _{\sigma\sigma '}f_{\sigma '}$.
In this case the operator $Q$ appearing in the numerator  
of Eq. (\ref{projection}) is not necessary to project away the $Q=0$ 
sector. In particular, the constrained $d$--electron Green's function is 
given in terms of the grand--canonical one ($G_{d}(\omega ,T,\lambda )$) 
as
\begin{equation} 
G_d(\omega ) = \lim _{\lambda \rightarrow \infty}
\frac {G_{d}(\omega ,T,\lambda )}{\langle Q \rangle _{G} (\lambda )} 
\label{eq:GdNCA}
\end{equation}
In the enlarged Hilbert space ($Q=0,1,2,...$) $G_{d}(\omega, T, \lambda )$ 
may be expressed in terms of the grand--canonical
pseudo--fermion and slave boson Green's functions 
using Wick's theorem. These auxiliary particle Green's functions, which
constitute the basic building blocks of the theory, are defined  
in imaginary time representation as

\begin{eqnarray}
{\cal G}_{f\sigma}(\tau_1-\tau_2) &=& - \langle \hat T \{f_\sigma
(\tau_1)f_\sigma^{\dag}(\tau_2)\}\rangle_G \\
{\cal G}_{b\bar\mu}(\tau_1 - \tau_2) &=& - \langle \hat T\{ b_{\bar\mu} (\tau_1)
b_{\bar\mu}^{\dag}(\tau_2)\}\rangle_G\ ,
\label{eq:def_gfb}
\end{eqnarray}
where $\hat T$ is the time ordering operator.  The Fourier transforms of
${\cal G}_{f,b}$ may be expressed in terms of the exact self--energies
$\Sigma_{f,b}$ as
\begin{equation}
{\cal G}_{f,b}(i\omega_n) = \Big\{[{\cal G}_{f,b}^0(i\omega_n)]^{-1} -
\Sigma_{f,b}(i\omega_n)\Big\}^{-1}
\label{green}
\end{equation}
where
\begin{eqnarray}
{\cal G}_{f\sigma}^0(i\omega_n) &=& (i\omega_n - E_d - \lambda)^{-1}\\
{\cal G}_{b\bar\mu}^0(i\omega_n) &=& (i\omega_n - \lambda)^{-1} 
\label{green0}
\end{eqnarray}
Since, as a consequence of the projection 
procedure $\lambda \rightarrow \infty$, the energy eigenvalues 
of $H + \lambda Q$  scale to infinity as $\lambda Q$, 
it is useful to shift the zero of the auxiliary particle
frequency scale by $\lambda$ (in the $Q = 1$ sector) and to define 
the ``projected'' Green's functions as 
\begin{equation}
G_{f,b}(\omega ) = \lim_{\lambda\to\infty} {\cal G}_{f,b}(\omega + \lambda)\ .
\label{eq:def_gfbproj}
\end{equation}
Note that this does not affect the energy 
scale of physical quantities (like the local $d$ electron Green's 
function), which is the {\it difference} between the 
the pseudo--fermion and the slave--boson energy.

{\em Canonical expectation values of conduction electron 
operators:}\/--- 
The canonical (i.e. projected onto the $Q=1$ subspace),
local conduction electron Green function is given as
\begin{equation}
G_{c\mu\sigma}(i\omega_n) = \Big\{[G_{c\mu\sigma}^0(i\omega_n)]^{-1} -
\Sigma_{c\mu\sigma}(i\omega_n)\Big\}^{-1}
\label{eq:greenc}
\end{equation}
with 
\begin{equation}
G_{c\mu\sigma}^0 (i\omega_n) = \sum _{\vec k} G^0_{c\mu\sigma}
(\vec k,i\omega_n)=
\sum _{\vec k} (i\omega_n +\mu _c - \epsilon_{\vec k}) ^{-1}\ ,
\label{eq:greenc0}
\end{equation}
where $\mu _c$ is the chemical potential of the conduction electrons.
The canonical, local $c$--electron self--energy,
$\Sigma_{c\mu\sigma} (i\omega _n)$, cannot be obtained from the
grand--canonical one by simply taking the limit $\lambda \rightarrow 
\infty$, since the $c$-electron density has a non--vanishing expectation 
value in the $Q=0$ subspace. However, it follows immediately from
the Anderson Hamiltonian, Eqs.~(\ref{2.5})--(\ref{2.6}), 
that the exact, canonical
conduction electron t--matrix $t_{\sigma\mu}(i\omega )$, defined by
$G_c = G^0_c [ 1+tG^0_c ]$,    
is proportional to the full, projected $d$--electron Green's function,
$t_{\mu\sigma}(i\omega ) = |V|^2G_{d\mu\sigma}$.
Thus, we have as an exact relation,
\begin{equation}
G_{c\mu\sigma} (i\omega_n) =
G_{c\mu\sigma}^0 (i\omega_n) \Big[ 1 +  |V|^2 G_{d\mu\sigma} 
(i\omega_n) G_{c\mu\sigma}^0 (i\omega_n) \Big] \  ,
\end{equation}
and by comparison with Eq. (\ref{eq:greenc}) 
we obtain the local conduction electron self--energy respecting the
constrained dynamics in the impurity orbital,
\begin{equation}
\Sigma_{c\mu\sigma} (i\omega_n) = \frac {|V|^2 G_{d\mu\sigma} 
(i\omega_n)}
{1+|V|^2 G_{c\mu\sigma}^0 (i\omega_n) G_{d\mu\sigma} 
(i\omega_n)}\ . 
\label{sigmac}
\end{equation}

Using phenomenological Fermi liquid theory \cite{Nozieres74} and 
also by means of perturbation theory to infinite order in 
the on--site repulsion $U$ \cite{yamada.75} it has been shown for 
the Fermi liquid case $M=1$ of the symmetric Anderson model 
($2E_d = -U$) that the exact $d$--electron
propagator $G_{d\sigma}(\omega )$ and the $d$--electron self--energy
$\Sigma _{d\sigma}(\omega )\equiv \omega - G_{d\sigma}(\omega )^{-1}$ 
obey the following local Fermi liquid relations 
in the limit $\omega \to 0-i0$, $T \to 0$,
\widetext
\top{-2.8cm}
\begin{eqnarray}
\mbox{Luttinger theorem:\ \ }&&
\int  d\omega\; f(\omega )
\frac{\partial \Sigma _{d\sigma}(\omega )}{\partial \omega}\;
G_{d\sigma }(\omega )=0
\label{luttinger}\\ \nonumber\\
\mbox{Friedel--Langreth:\ \ }&&
\frac{1}{\pi}{\rm Im} G_{d\sigma }(\omega ) = \frac{1}{\Gamma} \sin ^2
\Big(\frac{\pi n_d}{N}\Big) - c\; 
\Big[\Big(\frac{\omega}{T_K}\Big)^2+\Big(\frac{\pi T}{T_K}\Big)^2\Big]
\label{friedel}\\   \nonumber\\
&&{\rm Im} \Sigma _{d\sigma}(\omega ) = 
\frac{\Gamma}{\sin ^2(\pi n_d/N)} +
c \Big(\frac{\Gamma}{\sin ^2(\pi n_d/N)}\Big)^2  
\Big[\Big(\frac{\omega}{T_K}\Big)^2+\Big(\frac{\pi T}{T_K}\Big)^2\Big],
\label{friedel_sigma} \\ \nonumber 
\end{eqnarray}
\bottom{-2.7cm}
\narrowtext
where $c$ is a constant of $O(1)$.
Combining Eqs.~(\ref{sigmac}), (\ref{friedel}) it follows that 
(for $M=1$, away from particle hole symmetry) 
$\Sigma _{c\sigma}$ exhibits (in an exact theory) 
local Fermi liquid behavior as well, 
$\mbox{Im}\Sigma_{c\sigma}(\omega -i0,T=0) =a + b (\omega /T_K)^2$ 
for $\omega\rightarrow 0$. Note that this quantity is different from
the grand--canonical conduction electron self--energy and has a 
finite imaginary part at the Fermi level.

The momentum dependent conduction electron Green's function in the 
presence of a single impurity is given in terms of the canonical 
$d$--electron propagator as
\widetext
\top{-2.8cm}
\begin{eqnarray}
G_{c\mu\sigma}(\vec k,\vec k{'}\;;\;i\omega_n) G_{c\mu\sigma}^0
(\vec k,i\omega_n) \Big[\delta_{\vec k,\vec k{'}} + |V|^2 G_d(i\omega_n)
G_{c\mu\sigma}^0(\vec k{'},i\omega_n)\Big]\ . 
\end{eqnarray}
The latter expression is the starting point for treating a random
system of many Anderson impurities \cite{coxwilkins.87}.
\narrowtext
\subsection{Analytical properties and infrared behavior}
The Green's functions $G_{f,b,c}$ have the following spectral 
representations
\begin{equation}
G_{f,b,c}(i\omega_n) = \int_{-\infty}^\infty d\omega{'}
\frac{A_{f,b,c}(\omega{'})}{i\omega_n - \omega{'}}
\end{equation}
with the normalization of the spectral functions $A_{f,b,c}$
\begin{equation}
\int_{-\infty}^\infty d\omega A_{f,b,c}(\omega) = 1.
\end{equation}
Taking the limit $\lambda \rightarrow \infty$ has important 
consequences on the analytical structure of the auxiliary particle 
Green's functions:

(1) ---
It follows directly from the definitions Eqs. (\ref{eq:def_gfbproj}),
(\ref{eq:def_gfb}), using Eqs. (\ref{eq:stat_op}), (\ref{eq:exp_val}),
that the traces appearing in the canonical functions $G_{f,b}$ 
are taken purely over the $Q=0$ sector of Fock space 
\cite{note1}.
Thus, the backward--in--time ($\tau _1 < \tau _2$) or hole--like 
contribution to the auxiliary propagators in 
Eq. (\ref{eq:def_gfb}) vanishes after projection, and we have
\begin{eqnarray}
G_{f\sigma}(\tau_1-\tau_2) &=& -\Theta (\tau_1-\tau_2)
\lim _{\lambda\to \infty}
\langle f_{\sigma} (\tau_1)f^{\dag}_{\sigma} (\tau_2)\rangle _G\ 
\label{eq:def_gf1}\\
G_{b\bar\mu}(\tau_1-\tau_2) &=& -\Theta (\tau_1-\tau_2)
\lim _{\lambda\to \infty}
\langle b_{\bar\mu} (\tau_1)b^{\dag}_{\bar\mu} (\tau_2)\rangle _G\ .
\label{eq:def_gb1}
\end{eqnarray}
Consequently, their spectral functions $A_{f,b}$ have the Lehmann 
representation
\widetext
\top{-2.8cm}
\begin{eqnarray}
A_{f\sigma} (\omega )& = &\sum _{m,n\geq 0} \mbox{e}^{-\beta E_m^0}
\mid \langle 1,n | f_{\sigma}^{\dag} | 0,m\rangle \mid^2
\delta (\omega -( E_n^1 - E_m^0 )) 
\label{lehmannf}\\
A_{b\bar\mu} (\omega )& = &\sum _{m,n\geq 0} \mbox{e}^{-\beta E_m^0}
\mid \langle 1,n | b_{\bar\mu}^{\dag} | 0, m\rangle \mid^2
\delta (\omega -(E_n^1 - E_m^0 ))\ ,
\label{lehmannb}
\label{lehmann}
\end{eqnarray}
\bottom{-2.7cm}
\narrowtext
where $E_n^Q$ are the
energy eigenvalues ($E_0^0 \leq E_n^Q$ is the ground state energy) and 
$|Q,n\rangle $ the many--body eigenstates of $H$ in the sector $Q$ of 
Fock space. At zero temperature, $A_f$ reduces to
$A_{f\sigma} (\omega ) = \sum _{n\geq 0} \mid 
\langle 1,n | f_{\sigma}^{\dag} | 0,0\rangle \mid^2
\delta (\omega -( E_n^1 - E_0^0 ))$ and similar for $A_b$. 
It is seen that the $A_{f,b}$ have threshold behavior at 
$\omega = E_0 \equiv E_0^1 - E_0^0$, 
with $A_{f,b}(\omega ) \equiv 0$ for $\omega < E_0$, $T=0$. 
The vanishing imaginary part at frequencies $\omega < 0$ may be
shown to be a general property of all quantities involving slave particle
operators, e.g. also of auxiliary particle self--energies and
vertex functions.

(2) --- As will be seen in section V B (Eq. (\ref{gdNCA})), 
physical expectation values not only involve the particle--like
auxiliary propagators Eqs.~(\ref{eq:def_gf1}), (\ref{eq:def_gb1})
but also hole--like
contributions. It is, therefore, useful to define the ``anti--fermion'' 
and ``anti--boson'' propagators (in imaginary time representation)
\begin{eqnarray}
G^-_{f\sigma}(\tau_1-\tau_2) &=& - \Theta (\tau_2-\tau_1)\lim 
_{\lambda\to \infty}
\langle f^{\dag}_{\sigma} (\tau_2) f_{\sigma} (\tau_1) \rangle _G\ \\
G^-_{b\bar\mu}(\tau_1-\tau_2) &=& - \Theta (\tau_2-\tau_1)\lim 
_{\lambda\to \infty}
\langle b^{\dag}_{\bar\mu} (\tau_2) b_{\bar\mu} (\tau_1)\rangle _G\ ,
\end{eqnarray}
whose spectral functions have the Lehmann representations
\widetext
\top{-2.8cm}
\begin{eqnarray}
A^{-}_{f\sigma} (\omega )&=&\sum _{m,n\geq 0} \mbox{e}^{-\beta E_m^1}
\mid \langle 0,n | f_{\sigma} |
1,m\rangle \mid^2\delta (\omega -( E_n^0 - E_m^1)) 
\label{lehmannf-}\\
A^{-}_{b\bar\mu} (\omega )&=&\sum _{m,n \geq 0} \mbox{e}^{-\beta E_m^1}
\mid \langle 0,n | b_{\bar\mu} |
1,m\rangle \mid^2\delta (\omega - ( E_n^0 - E_m^1 ))\ .
\label{lehmannb-}
\label{lehmann-}
\end{eqnarray}
\bottom{-2.7cm}
\narrowtext
$E_0^1$ is the ground state energy in the $Q=1$ sector. The
expressions (\ref{lehmannf}), (\ref{lehmannb}) and 
(\ref{lehmannf-}), (\ref{lehmannb-}) immediately imply
a relation between $A_{f,b}$ and $A^-_{f,b}$,
\begin{equation}
A^-_{f,b} (\omega ) = \mbox{e}^{-\beta \omega } A_{f,b}(\omega ) \ .
\label{Aminus}
\end{equation}

(3) --- The property of only forward--in--time propagation 
(Eqs. (\ref{eq:def_gf1}), (\ref{eq:def_gb1})) means
that the auxiliary particle propagators $G_{f,b}$
are formally identical to the core hole propagators of the well--known
X--ray threshold problem \cite{anderson.67}--\cite{mahan.80}. Thus, the 
knowledge of the infrared behavior of the latter may be directly 
applied to the former. In particular, the spectral functions are 
found (see below) to diverge at the threshold $E_0$ in  
a power law fashion (infrared singularity)
\begin{equation}
A_{f,b}(\omega ) \sim | \omega - E_0 |
^{-\alpha_{f,b}}\theta (\omega - E_0)
\label{powerlaw}
\end{equation}
due to a diverging number of particle--hole excitation processes in the
conduction electron sea as $\omega \to E_o$. 
For the single channel case $(M=1)$, i.e. the usual Kondo or mixed
valence problem, the exponents $\alpha_f$ and $\alpha_b$ can be found
analytically from the following chain of arguments:  Anticipating that 
in this case the impurity spin is completely screened by the conduction
electrons at temperature $T = 0$, leaving a pure-potential scattering
center, the ground state $| 1,0 \rangle$ is a Slater determinant of 
one--particle scattering states, characterized by scattering phase 
shifts $\eta_\sigma$ in the s--wave channel (assuming for simplicity a
momentum independent hybridization matrix element $V$).
To calculate the fermion spectral function $A_{f\sigma}(\omega)$ at $T=0$ 
from Eq.~(\ref{lehmannf}), one needs to 
evaluate $\langle 1,n\mid f_\sigma^{\dag}\mid 0,0 \rangle$, which
is just the overlap of two slater determinants, an eigenstate of the 
fully interacting Kondo system, $|1,n\rangle$, on the one hand, and the 
ground state of the conduction electron system in the absence of the
impurity combined with the decoupled impurity level occupied by an electron
with spin $\sigma$, $f^{\dag}_{\sigma}|0,0\rangle$, on the other hand.  
As shown by Anderson \cite{anderson.67}, 
the overlap of the two ground state slater determinants,
$\langle 1,0\mid f_\sigma^{\dag}\mid 0,0 \rangle$, tends to zero in
the thermodynamic limit (orthogonality catastrophe). 
Analogous relations hold for the boson spectral function 
$A_{b}(\omega)$. As a result, the exponential relaxation into the 
interacting ground state for long times is inhibited, 
leading to the infrared power 
law divergence of the spectral functions, Eq. (\ref{powerlaw}).

The X--ray threshold exponents can be expressed in terms of the 
scattering phase shifts at the Fermi level by the exact relation 
\cite{schotte.69}
\begin{equation}
\alpha_{f,b} = 
1 - \sum_{\sigma '}\Big(\frac{\eta_{f,b\;\sigma '}}{\pi}\Big)^2\ .
\end{equation}
Here the $\eta_{f\sigma '}$ ($\eta_{b\sigma '}$) are the scattering 
phase shifts of the single--particle wave functions in channel $\sigma '$
of the fully interacting ground state $| 1,0\rangle$,
relative to the wave functions of the free state $f_\sigma^{\dag}| 
0,0\rangle$ (\,$b^{\dag}| 0,0\rangle$\,). 
Via the Friedel sum rule, the scattering phase shifts are, in turn, 
related to the change $\Delta n_{c\sigma '}$ of the average number of 
conduction electrons per scattering channel $\sigma '$ due to the 
presence of the impurity: 
$\eta_{f,b\; \sigma '}=  \pi \Delta n_{c\sigma '}$. Obviously, 
$\Delta n_{c\sigma '}$ is equal and opposite in sign to the difference 
of the average impurity occupation numbers of the states  $| 1,0\rangle$ 
and $f_\sigma^{\dag}| 0,0\rangle$ (\,$b^{\dag}| 0,0\rangle$\,). 
Thus, in the pseudofermion propagator $G_{f\sigma }$ we have the 
phase shifts,
\begin{equation}
\eta_{f\sigma '}= - \pi \Big(\frac{n_d}{N}-\delta _{\sigma\sigma '}\Big) 
\end{equation}
and in the slave boson propagator $G_b$,
\begin{equation}
\eta_{b}= - \pi\frac{n_d}{N}\ , 
\end{equation}
where $n_d$ denotes the total occupation number of the impurity level 
in the interacting ground state. (The term $\delta _{\sigma\sigma '}$ in
$\eta_{f\; \sigma '}$ appears because  $f_\sigma^{\dag}| 0,0\rangle$ 
has impurity occupation number 1 in the channel $\sigma$.)
For example, in the Kondo limit $n_d \rightarrow 1$ and for a 
spin $1/2$ impurity 
($N=2$) this leads to resonance scattering, $\eta_{f,b\;\sigma '} = 
\pi/2$. As a result, one finds \cite{mengemuha.88} for the threshold 
exponents
\begin{eqnarray}
\alpha_f &=& \frac{2n_d - n_d^2}{N}\\
\alpha_b &=& 1-\frac{n_d^2}{N}
\end{eqnarray}
These results have been found independently from Wilson's numerical
renormalization group approach \cite{costi.94,costi.94b} and using the Bethe 
ansatz solution and boundary conformal field theory \cite{fujimoto.96}.  
It is interesting to note that (i) the exponents depend on the 
level occupancy $n_d$ 
(in the Kondo limit $n_d \rightarrow 1$, $\alpha_f = 1/N$ and
$\alpha_b = 1 - 1/N$, whereas in the opposite, empty orbital, limit
$n_d \rightarrow 0$, $\alpha_f \rightarrow 0$ and $\alpha_b
\rightarrow 1$) (ii) the sum of the exponents $\alpha_f + \alpha_b = 1
+ 2 \frac{n_d(1-n_d)}{N} \geq  1$.

We stress that the above derivation of the infrared exponents
$\alpha_{f,b}$ holds true only if the impurity complex acts as a pure
potential scattering center at $T = 0$.  This is equivalent to the
statement that the conduction electrons behave locally, i.e. at the
impurity site, like a Fermi liquid. Conversely, in the multi--channel 
(non--FL) case, $N\geq 2$, $M\geq N$, the exponents have been found 
from a conformal field theory solution \cite{affleck.91} of the problem
in the Kondo limit to be 
\begin{eqnarray}
\alpha _f &=& M/(M+N)\nonumber \\
\alpha _b  &=& N/(M+N)
\label{40c}
\end{eqnarray}
which differ from the FL values. Thus, one may infer from
the values of $\alpha_{f,b}$ as a function of $n_d$, whether or not
the system is in a local Fermi liquid state.

\section{Mean field approach and $1/N$ expansion at $U\rightarrow \infty$}

For physical situations of interest, the $s-d$ hybridization of the
Anderson model (\ref{sbhamilton}) is much smaller than the conduction 
band width, ${\cal N}(0) V \ll 1$, where ${\cal N}(0) =1/D$ 
is the local conduction 
electron density of states at the Fermi level.  This suggests a 
perturbation expansion in ${\cal N}(0)V$. A straightforward expansion in terms 
of bare Green's functions is not adequate, as it would not allow to 
capture the physics of the Kondo screened state, or else the infrared
divergencies of the auxiliary particle spectral functions discussed in
the last section.  In the framework of the slave boson representation,
two types of nonperturbative approaches have been developed.  The
first one is mean field theory for both the slave boson amplitude 
$\langle b\rangle$ and the constraint 
($\langle Q\rangle = 1$ rather than $Q = 1$).  The second one
is resummation of the perturbation theory to infinite order.

\subsection{Slave boson mean field theory} 
Slave boson mean field theory is based on the assumption that the slave
bosons condense at low temperatures such that $\langle b_{\bar\mu}\rangle\neq 
0$.
Replacing the operator $b_{\bar\mu}$ in $H + \lambda Q$  
by $\langle b_{\bar\mu} \rangle$ (see Ref.~\cite{readnewns.88}), 
where $\lambda$ is a Lagrange multiplier to be adjusted such that
$\langle Q \rangle = 1$, one arrives at a resonance level model for the
pseudofermions.  The position of the resonance, $E_d + \lambda$, is
found to be given by the Kondo temperature $T_K$, and is thus close to
the Fermi energy. The resonance generates the low energy scale $T_K$,
and leads to local Fermi liquid behavior.  While this is qualitatively
correct in the single--channel case, it is in blatant disagreement with
the exactly known behavior in the multi--channel case.  The mean field
theory can be shown to be exact for $M=1$ in the limit $N \rightarrow
\infty$ for a model in which the constraint is softened to be $Q=N/2$. 
However, for finite $N$ the breaking of the local gauge symmetry,
which would be implied by the condensation of the slave boson field,
is forbidden by Elitzur's theorem \cite{elitzur.75}.
It is known that for finite $N$ the fluctuations 
in the phase of the complex expectation value $\langle b_{\bar\mu}\rangle$ 
are divergent and lead to the suppression of $\langle b_{\bar\mu} \rangle$ to 
zero (see also \cite{jevicki.77}--\cite{lawrie.85}). 
This is true in the cartesian
gauge, whereas in the radial gauge the phase fluctuations may be shown
to cancel at least in lowest order. It has not been possible
to connect the mean field solution, an apparently reasonable
description at low temperatures and for $M = 1$, to the high 
temperature behavior $(T \gg T_K)$, dominated by logarithmic temperature 
dependence, in a continuous way \cite{readnewns.88}.  
Therefore, it seems that the slave boson mean field
solution does not offer a good starting point even for only a
qualitatively correct description of quantum impurity models.

\subsection{1/N expansion vs. self--consistent formulation}
The critical judgement of mean field theory is corroborated by the
results of a straightforward $1/N$-expansion in the single channel
case, keeping the exact constraint, and not allowing for a finite
bose field expectation value \cite{kuroda.88}.  
Within this scheme the exact behavior of the thermodynamic quantities 
(known from the Bethe ansatz solution) at low temperatures as well as 
high temperatures is recovered to the considered order in $1/N$.  
Also, the exact auxiliary particle exponents $\alpha _{f,b}$ are 
reproduced in order $1/N$, using a plausible exponentiation 
scheme \cite{kuroda.97}.

In addition, dynamical quantities like the d-electron spectral
function and transport coefficients can be calculated exactly to a
desired order in $1/N$ within this approach.  However, as clear-cut
and economical this method may be, it does have serious limitations.
For once, the $1/N$ expansion is not uniformly convergent as a
function of temperature \cite{sellier98}.  Rather, the expansions at low temperature
and at high temperature have to be done around two different
saddle-points (the limits $N\rightarrow\infty$).  It is not known how
to match these expansions in the crossover region $T\sim T_K$ in a
systematic way.
Secondly, the experimentally most relevant case of $N=2$ or somewhat
larger is not accessible in $1/N$ expansion. Thirdly, non-Fermi
liquid behavior, being necessarily non-perturbative in $1/N$, cannot
be dealt with in a controlled way on the basis of a $1/N$-expansion.
To access these latter two regimes, a new approach non-perturbative 
in $1/N$ is necessary.

\section{Conserving approximations: gauge invariant self-consistent
perturbation theory in the hybridization}
\label{conserving}

We conjecture that this new approach is gauge invariant many-body
theory of pseudofermions and slave bosons. As long as gauge symmetry
violating objects such as Bose field expectation values or fermion
pair correlation functions do not appear in the theory, gauge
invariance of physical quantities can be guaranteed in suitably chosen
approximations by the proper match of pseudofermion and slave boson
properties, without introducing an additional gauge field.  This
requires the use of conserving approximations \cite{kadanoff.61}, 
derived from a Luttinger-Ward functional $\Phi$.

\subsection{Generating functional}
\label{functional}
  
$\Phi$ consists of all vacuum
skeleton diagrams built out of fully renormalized Green's functions
$G_{b,f,c}$ and the bare vertex $V$.  The self--energies
$\Sigma_{b,f,c}$ are obtained by taking the functional derivative of
$\Phi$ with respect to the corresponding Green's function (cutting the
Green's function line in each diagram in all possible ways),
\begin{equation}
\Sigma_{b,f,c} = \delta \Phi/\delta G_{b,f,c}.
\label{fderiv}
\end{equation}
Irreducible vertex functions, figuring as integral kernels in
two-particle Bethe-Salpeter equations, are generated by second order
derivatives of $\Phi$.

The choice of diagrams for $\Phi$ defines a given approximation. It
should be dictated by the dominant physical processes and by
expansion in a small parameter, if available.  Even in the presence of
a small parameter (in this case ${\cal N}_0V$), straightforward low order
renormalized perturbation theory may not give an even qualitatively
correct result, if singular vertex functions appear.  It is therefore
necessary to check whether vertex functions become singular at the
level where approximate single particle Green's functions are used to
calculate the integral kernels of the respective vertex functions.
Should this be the case, the vertex functions have to be included into
the approximation in a gauge invariant way.  As we shall see, in the
present case it is required to take all two-particle vertex functions
into consideration.  Vertex functions involving three or more
interacting particles will be omitted in the hope that the
corresponding phase space is small so that they will contribute less
even if they are singular.  This is the first fundamental
approximation.  The second one is that we will approximate the
irreducible kernels in the Bethe-Salpeter equations for the vertex
functions by the lowest order diagram in ${\cal N}_0V$.
\widetext
\begin{figure}
\vspace*{-0cm}
\centerline{\psfig{figure=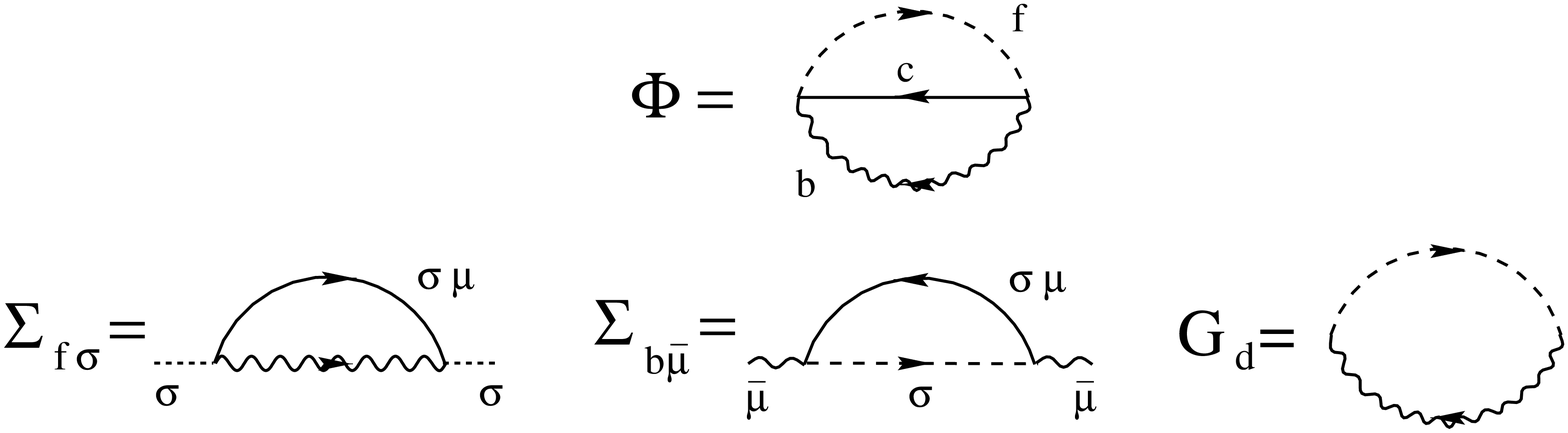,width=0.8\textwidth}}
\vspace*{0.6cm}
\caption{
Diagrammatic representation of the generating functional $\Phi $ 
of the NCA. Also shown are the pseudoparticle self--energies and the
local electron Green's function derived from $\Phi$, Eqs.~(19)--(21).
Throughout this article, dashed, wavy and solid lines represent fermion, 
boson, and conduction electron lines, respectively. In the diagram for 
$\Sigma _{f\sigma}$ the spin labels are shown explicitly to 
demonstrate that there are no coherent spin fluctuations taken into account.
\label{NCA}}
\end{figure}
\narrowtext
\subsection{Non-crossing approximation (NCA)}
\label{non-crossing}

  As noted before, in the
present context, we may take the hybridization $V$ to be a small
quantity (dimensionless parameter ${\cal N}_oV$).  This suggests to start
with the lowest order (in $V$) diagram of $\Phi$, which is second
order (see Fig.~\ref{NCA}).  The self--energies generated from this 
obey after analytic continuation to real frequencies 
($i\omega \rightarrow \omega -i0$) and projection the following 
equations of self--consistent second order perturbation theory 
\widetext
\top{-2.8cm}
\begin{eqnarray}
\Sigma_{f\sigma}^{(NCA)}(\omega-i0)&=&\Gamma\sum _{\mu}\int 
              \frac{{d}\varepsilon}{\pi}\,
               [1-f(\varepsilon )]
              A_{c\mu\sigma}^0(\varepsilon)G_{b\bar\mu}(\omega -\varepsilon -i0)
              \label{sigfNCA}\\
\Sigma_{b\bar\mu}^{(NCA)}(\omega -i0)&=&\Gamma\sum _{\sigma}\int 
              \frac{{d}\varepsilon}{\pi}\,
              f(\varepsilon )A_{c\mu\sigma}^0(\varepsilon)
              G_{f\sigma}(\omega +\varepsilon -i0)
              \label{sigbNCA}\\
G_{d\mu\sigma}^{(NCA)}(\omega -i0)
         &=& \int  {d}\varepsilon\,  {\rm e}^{-\beta\varepsilon}
         [ G_{f\sigma}(\omega +\varepsilon -i0)A_{b\bar\mu}(\varepsilon )
         -A_{f\sigma}(\varepsilon )
                   G_{b\bar\mu}(\varepsilon -\omega +i0) ]
              \label{gdNCA}                               \\
         &=& \int  {d}\varepsilon\,  
         [ G_{f\sigma}(\omega +\varepsilon -i0)A^-_{b\bar\mu}(\varepsilon )
          -A^-_{f\sigma}(\varepsilon )
                   G_{b\bar\mu}(\varepsilon -\omega +i0) ]\; , \nonumber
\end{eqnarray}
\bottom{-2.7cm}
\narrowtext
where $A_{c\mu\sigma}^0=\frac{1}{\pi}\, 
{\rm Im}G_{c\mu\sigma}^0/{\cal N}(0)$ is the
(free) conduction electron density of states per spin and channel,
normalized to the density of states at the Fermi level ${\cal N}(0)$, and 
$f(\varepsilon )=1/({\rm exp}(\beta\varepsilon )+1)$ denotes the 
Fermi distribution function. Together with the expressions 
(\ref{green}), (\ref{green0}) for the Green's functions,
Eqs. (\ref{sigfNCA})--(\ref{gdNCA}) form a set of self--consistent 
equations for $\Sigma_{b,f,c}$, comprised of all diagrams without 
any crossing propagator lines and are, thus, known as the
``non--crossing approximation'', in short NCA \cite{keiter.71,kuramoto.83}.

At zero temperature and for low frequencies, Eqs. (\ref{sigfNCA}) 
and (\ref{sigbNCA}) may be converted into a set of linear differential 
equations for $G_f$ and $G_b$ \cite{muha.84}, 
which allow to find the infrared exponents as $\alpha_f =
\frac{M}{M+N}$; $\alpha_b = \frac{N}{M+N}$, independent of $n_d$.  For
the single channel case these exponents do not agree with the exact
exponents derived in section III B. This indicates that the NCA is
not capable of recovering the local Fermi liquid behavior for $M=1$.
A numerical evaluation of the $d$-electron Green's function, which 
is given by the local self--energy $\Sigma_c$ divided by $V^2$ and 
hence is given by the boson--fermion bubble within NCA (Fig.~\ref{NCA}), 
shows indeed a spurious singularity at the Fermi energy \cite{costi.96}.
The NCA performs somewhat better in the multi--channel case, 
where the exponents $\alpha_f$ and $\alpha_b$ yield the correct 
non--Fermi liquid exponents
of physical quantities as known from the Bethe ansatz 
solution \cite{Wieg83} and conformal field 
theory \cite{affleck.91}. However, the specific heat and
the residual entropy are not given correctly in NCA.  Also, the
limiting low temperature scaling laws for the thermodynamic
quantities are attained only at temperatures substantially below
$T_K$, in disagreement with the exact Bethe ansatz solution.

\subsection{Evaluation of the self--consistency 
equations at low temperatures}
In order to enter the asymptotic power law regime of the auxiliary 
spectral functions, the self--consistent equations 
must be evaluated for temperatures several orders of magnitude below $T_K$, 
the low temperature scale of the model. The equations are solved 
numerically by iteration. In the following we describe the two main 
procedures to make the diagrammatic auxiliary particle technique 
suitable for the lowest temperatures.

The grand--canonical expectation value of the auxiliary particle 
number appearing in Eq.~(\ref{eq:GdNCA}) is given in terms of the 
grand--canonical (unprojected) auxiliary particle spectral functions 
${\cal A}_{f,b}(\omega,\lambda)$ by,
\begin{eqnarray}
&&\langle Q \rangle _{G} (\lambda ) = \\
&&\int  {d}\omega
\Big[ f(\omega) \sum _{\sigma } {\cal A}_{f\sigma } (\omega,\lambda)
+ b(\omega) \sum _{\mu } {\cal A}_{b\bar\mu}(\omega,\lambda)
\Big] \nonumber
\end{eqnarray}
where $f(\omega)$, $b(\omega)$ are the Fermi and Bose distribution 
functions, respectively. 
Substituting this into the expression (\ref{eq:zcan}) for the 
canonical partition function we obtain after carrying out
the transformation $\omega \rightarrow \omega +\lambda $, and taking
the limit $\lambda \rightarrow \infty $
\begin{eqnarray}
e^{-\beta F_{imp}(T)} & \equiv & 
\frac{Z_{C}}{Z_{Q=0}}  = 
\lim_{\lambda\rightarrow\infty}e^{\beta\lambda}
\langle Q \rangle_{G}(\lambda) \label{eq:Qproj}\\
& = & \int {d}\omega e^{-\beta\omega} \Big[  \sum _{\sigma } 
A_{f\sigma }(\omega) +\sum _{\mu} A_{b\bar\mu}(\omega)\Big]\ .
\nonumber
\end{eqnarray}
By definition 
$F_{imp}=-\frac{1}{\beta}\ln(Z_{C}/Z_{Q=0})$ is the impurity 
contribution to the Free energy.

The numerical evaluation of expectation values like 
$\langle Q\rangle _{G}(\lambda\rightarrow\infty)$ (Eq. (\ref{eq:Qproj}))
or $G_{d\mu\sigma}(\omega,\lambda\rightarrow\infty)$ (Eq. (\ref{gdNCA})) 
is non--trivial, (1) because at $T=0$ the auxiliary spectral functions 
$A_{f,b}(\omega,T)$
are divergent at the threshold frequency $E_{0}$, where the exact
position of $E_{0}$ is a priori not known, and (2) because 
the Boltzmann factors $e^{-\beta\omega}$ diverge exponentially for 
$\omega < 0$. Therefore, we apply the following transformations:

(1) Before performing the projection $\omega \rightarrow \omega +\lambda$,
$\lambda\rightarrow\infty$ we re--define the frequency scale of all
auxiliary particle functions $A_{f,b}$ according to  
$\omega \rightarrow \omega +\lambda_0$, where $\lambda_0$ is a finite
parameter. In each iteration $\lambda _0$  is then determined  such that
\begin{equation}
\int {d}\omega
e^{-\beta\omega} \Big[ \sum _{\sigma } A_{f\sigma }(\omega) +
\sum _{\mu} A_{b\bar\mu}(\omega)\Big]=1
\label{eq:fixl}
\end{equation}
where $A_{f,b}(\omega)=\lim_{\lambda\rightarrow\infty}
A_{f,b}(\omega+\lambda_0,\lambda)$ 
is now an auxiliary spectral function with the new reference energy.
It is seen by comparison with Eq.~(\ref{eq:Qproj}) that 
$\lambda_0(T)=F_{imp}(T) = F_{Q=1}(T)-F_{Q=0}(T)$, i.e. $\lambda_0$ 
is the chemical potential for the auxiliary particle number $Q$, or 
equivalently the impurity contribution to the Free energy. 
The difference of the Free energies becomes equal to the threshold energy 
$E_{0}=E_{Q=1}^{GS}-E_{Q=0}^{GS}$ at $T=0$. More importantly, however, 
the above way of determining a ``threshold'' is less {\em ad hoc}\/ than,
for example, defining it by a maximum in some function appearing in the 
NCA equations. It is also seen from Eq.~(\ref{eq:fixl}) that this procedure 
defines the frequency scale of the auxiliary particles such that 
the $T=0$ threshold divergence of the
spectral functions is at the {\it fixed} frequency $\omega = 0$.
This substantially increases the precision as well as the speed of 
numerical evaluations. Eq. (\ref{eq:GdNCA}) for the projected 
$d$--electron Green's function becomes
\begin{equation}
G_d(\omega ) = \lim _{\lambda \rightarrow \infty} e^{\beta\lambda}
G_{d}(\omega,T,\lambda ).
\end{equation}
  
(2) The divergence of the Boltzmann factors implies that the 
self--con\-sis\-tent solutions for $A_{f,b} (\omega)$
vanish exponentially $\sim e^{\beta\omega}$ for negative frequencies,
confirming their threshold behavior.
It is convenient, not to formulate the self--consistent equations in 
terms of $A_{f,b}$ like in earlier evaluations \cite{bickers.87}, 
but to define new functions $\tilde A_{f,b}(\omega)$ and 
$\mbox{Im}\tilde\Sigma_{f,b}(\omega)$ such that
\begin{eqnarray}
A_{f,b} (\omega) & = & f(-\omega )~ \tilde A_{f,b}(\omega)\\
\mbox{Im}\Sigma_{f,b}(\omega) & = & f(-\omega ) ~ \mbox{Im} 
\tilde\Sigma_{f,b}(\omega).
\end{eqnarray}
After fixing the chemical potential $\lambda _0$ and performing the
projection onto the physical subspace, the canonical partition function
(Eq.~(\ref{eq:zcan})) behaves as 
$\lim _{\lambda\rightarrow\infty}e^{\beta(\lambda-
\lambda_0)}\;Z_C(T) =1$, and from Eq.~(\ref{Aminus}) we have
$A_{f,b}^-(\omega ) = f(\omega ) \tilde A _{f,b} (\omega )$. 
In this way all exponential divergencies are absorbed by one single 
function for each particle species.
As an example, the NCA equations in terms of these functions are 
free of divergencies of the statistical factors and read
\widetext
\top{-2.8cm}
\begin{eqnarray}
\mbox{Im}\tilde\Sigma_{f\sigma}(\omega -i0) 
& = & \Gamma\sum_{\mu} \int {d}\varepsilon\,
\frac{f(-\varepsilon)(1-f(\omega -\varepsilon))}{1-f(\omega )} 
A_{c\mu\sigma}^0(\varepsilon)
\tilde A_{b\bar\mu}(\omega-\varepsilon)\\
\mbox{Im}\tilde\Sigma_{b\bar\mu}(\omega -i0) 
& = & \Gamma\sum_{\sigma} \int {d}\varepsilon\,
\frac{f(\varepsilon)(1-f(\omega +\varepsilon))}{1-f(\omega )} 
A_{c\mu\sigma}^0(\varepsilon)
\tilde A_{f\sigma}(\omega+\varepsilon)\\
\langle Q \rangle  (\lambda _0,\lambda\rightarrow\infty) & = &
\int {d}\omega
f(\omega)\Big[ \sum _{\sigma } \tilde A_{f\sigma }(\omega) +
\sum _{\mu}\tilde A_{b\bar\mu} (\omega)\Big] = 1\\
\mbox{Im}G_{d\sigma}(\omega -i0) & = & 
\int {d}\varepsilon [f(\varepsilon +\omega)f(-\varepsilon )+
f(-\varepsilon -\omega)f(\varepsilon)]
\tilde A_{f\sigma}(\varepsilon +\omega)\tilde 
A_b(\varepsilon) .\\ \nonumber
\end{eqnarray}
\begin{figure}
\vspace*{-0cm}
\centerline{\psfig{figure=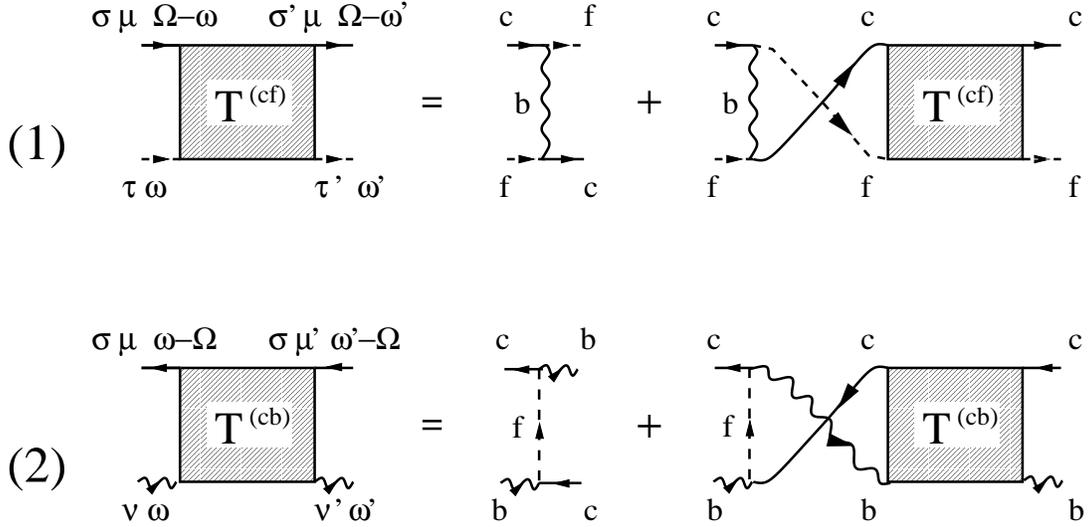,width=0.8\textwidth}}
\vspace*{0.6cm}
\caption{
Diagrammatic representation of the Bethe--Salpeter equation for (1) 
the conduction electron--pseudofermion
T-matrix $T^{(cf)}$, Eq.~(\ref{cftmateq}), and (2) the conduction 
electron--slave boson T-matrix $T^{(cb)}$, Eq.~(\ref{cbtmateq}). 
$T^{(cb)}$ is obtained from $T^{(cf)}$ by 
interchanging $f \leftrightarrow b$ and $c\leftrightarrow c^{\dag}$.
\label{tmat}}
\end{figure}
\narrowtext
The real parts of the self--energies $\Sigma _f$, $\Sigma _b$ 
are determined from $\mbox{Im}\Sigma _f$, $\mbox{Im}\Sigma _b$ 
through a Kramers--Kroenig relation, and the auxiliary functions
$\tilde A_{f\sigma}(\omega)=\frac{1}{\pi}
\mbox{Im}\tilde\Sigma _{f\sigma}(\omega -i0)/
\bigl[\bigl(\omega+\lambda_0-i0-E_d-\mbox{Re}\Sigma_{f\sigma}
(\omega-i0)\bigr)^2+\mbox{Im}\Sigma_{f\sigma}(\omega-i0)^2\bigr]$, 
$\tilde A_{b\bar\mu}(\omega)=\frac{1}{\pi}
\mbox{Im}\tilde\Sigma _{b\bar\mu}(\omega -i0)/
\bigl[\bigl(\omega+\lambda_0-i0-\mbox{Re}\Sigma_{b\bar\mu}
(\omega-i0)\bigr)^2+\mbox{Im}\Sigma_{b\bar\mu}(\omega-i0)^2\bigr]$,
thus closing the above set of equations. 

The method described above allows to solve the NCA equations 
effectively for temperatures down to typically $T=10^{-4}T_K$. 
It may be shown that the procedures described above can also be applied to 
self--consistently compute vertex corrections beyond the NCA 
(see section VI), thus avoiding any divergent statistical factors
in the selfconsistency equations.

\section{Conserving T-matrix approximation (CTMA) at $U\rightarrow \infty$}

\subsection{Dominant contributions at low energy}
In order to eliminate the shortcomings of the NCA mentioned above, 
the guiding principle should be to find 
contributions to the vertex functions which renormalize the
auxiliary particle threshold exponents to their correct values,
since this is a necessary condition for the description of 
FL and non--FL behavior, as discussed in section III B. Furthermore,
it is instructive to realize that in NCA any coherent spin flip
and charge transfer processes are neglected, as can be seen 
explicitly from Eqs.~(\ref{sigfNCA}), (\ref{sigbNCA}) or from 
Fig.~\ref{NCA}. These processes are known to be responsible for the
quantum coherent collective behavior of the Anderson impurity
complex below $T_K$. The existence of collective excitations
in general is reflected in a singular behavior of the 
corresponding two--particle vertex functions. In view of the
tendency of Kondo systems to form a collective spin singlet
state, we focus our attention on the two-particle vertex functions, in
particular, in the spin singlet channel
of the pseudofermion--conduction electron vertex function and
in the slave boson--conduction electron vertex function.
It may be shown by power counting arguments (compare appendix A)
that there are no corrections to the NCA exponents in any finite 
order of perturbation theory \cite{coxruck.93}. Thus, we 
are led to search for singularities in the aforementioned vertex 
functions arising from an infinite resummation of terms.   

From the preceding discussion it is natural to perform a partial 
resummation of those terms which, at each order in the 
hybridization $V$, contain the maximum number of spin flip or charge 
fluctuation processes, respectively. This amounts to calculating the 
conduction electron--pseudofermion vertex function in
the ``ladder'' approximation defined in Fig.~\ref{tmat}, where the 
irreducible vertex is given by $V^2G_b$. In analogy to similar 
resummations for an interacting one--component Fermi system, we 
call the total c--f vertex function T--matrix $T^{(cf)}$. 
The Bethe--Salpeter equation for $T^{(cf)}$ reads (Fig.~\ref{tmat} (1)),
\widetext
\top{-2.8cm}
\begin{eqnarray}
T^{(cf)\ \mu}_{\sigma\tau,\sigma '\tau '}
(i\omega _n, i\omega _n ', i\Omega _n ) =
&+&V^2G_{b\bar\mu}(i\omega _n + i\omega _n ' - i\Omega _n ) 
\delta _{\sigma\tau '}\delta _{\tau \sigma '}
\nonumber\\
&-&V^2T\sum _{\omega _n''}G_{b\bar\mu}(i\omega _n + i\omega _n '' - 
i\Omega _n ) 
G_{f\sigma}(i\omega _n'') \ G^0_{c\mu\tau}(i\Omega _n -i\omega _n '')\ 
T^{(cf)\ \mu}_{\tau \sigma,\sigma '\tau '}(i\omega _n '', i\omega _n ', 
i\Omega _n ). 
\label{cftmateq}
\end{eqnarray}
\noindent
A similar  integral equation holds for the charge fluctuation T--matrix
$T^{(cb)}$ (Fig.~\ref{tmat} (2)), 
\begin{eqnarray}
T^{(cb)\ \sigma}_{\mu\nu,\mu '\nu '}(i\omega _n,i\omega _n ',i\Omega _n) =
&+&V^2G_{f\sigma}(+i\omega _n + i\omega _n ' - i\Omega _n ) 
\delta _{\mu\nu '}\delta _{\nu \mu ' }
\nonumber\\
&-&V^2T\sum _{\omega _n''}G_{f\sigma}(i\omega _n + i\omega _n '' - 
i\Omega _n ) 
G_{b\bar\mu}(i\omega _n'') \ G^0_{c\nu\sigma}(-i\omega _n ''-i\Omega _n )\ 
T^{(cb)\ \sigma}_{\nu\mu,\mu '\nu '}(i\omega _n '', i\omega _n ', 
i\Omega _n ). 
\label{cbtmateq}
\end{eqnarray}
\bottom{-2.7cm}
\narrowtext
\noindent
In the above Bethe--Salpeter equations $\sigma$, $\tau$, $\sigma '$, 
$\tau '$ represent spin and $\mu$, $\nu$, $\mu '$, $\nu '$ channel
indices.  We note that these are the only two-particle vertex
functions after projection.  The principal approximation adopted here
is the form of the irreducible kernel, which we approximate
by the lowest order diagram.

\begin{figure}
\vspace*{-0cm}
\centerline{\psfig{figure=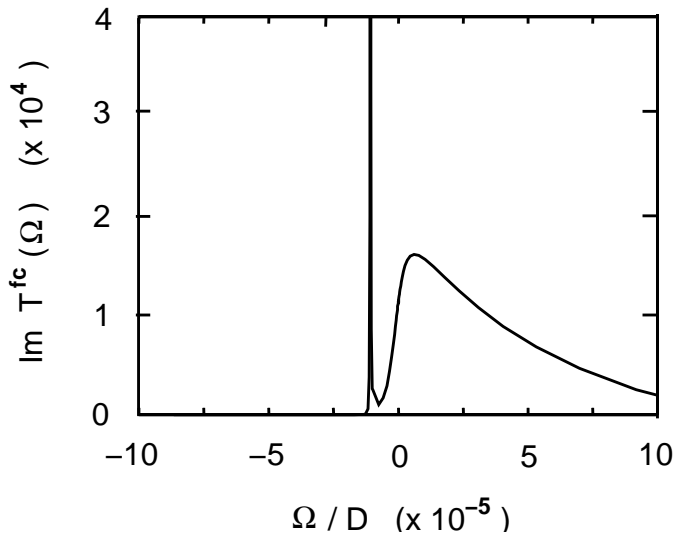,width=0.5\textwidth}}
\vspace*{0.6cm}
\caption{ 
Imaginary part of the conduction electron--pseudofermion 
T--matrix $T^{(cf)}$ as a function of the
COM frequency $\Omega$ for the single--channel case $M=1$, $N=2$, 
evaluated by inserting NCA solutions for the
intermediate state propagators ($E_d=-0.67 D$, $\Gamma = 0.15 D$,
$T=4\cdot 10^{-3}T_K$).
The contribution from the pole positioned at a negative frequency
$\Omega = \Omega _{cf} \simeq -T_K$ (compare text) is clearly seen.
\label{tmatpole0}}
\end{figure}

Inserting NCA Green's functions for the intermediate state
propagators of Eq. (\ref{cftmateq}) and solving it numerically, 
we find at low temperatures and in the Kondo regime $(n_d \gsim  0.7)$ 
a pole of $T^{(cf)}$ in the singlet channel (see appendix A) 
as a function of the center--of--mass (COM) frequency $\Omega$, at a 
frequency which scales with the Kondo temperature, 
$\Omega = \Omega_{cf} \simeq - T_K$. 
This is shown in Fig.~\ref{tmatpole0}. The threshold behavior of
the imaginary part of $T^{(cf)}$ as a function of $\Omega$ with
vanishing spectral weight at negative frequencies and temperature 
$T=0$ is clearly seen. In addition, a very sharp structure appears,
whose broadening is found to vanish as the temperature tends to zero, 
indicative of a pole in $T^{(cf)}$ at the {\it real} frequency
$\Omega _{cf}$, i.e.~the tendency to form a collective singlet 
state between the conduction electrons and the localized spin.   
\noindent
Similarly, the corresponding $T$-matrix $T^{(cb)}$ in the conduction
electron--slave boson channel, evaluated within the analogous 
approximation, develops a pole at negative values of $\Omega$ in the 
empty orbital regime $(n_d \lsim 0.3)$.  In the mixed valence 
regime ($n_d \simeq 0.5)$ the poles in both $T^{(cf)}$ 
and $T^{(cb)}$ coexist. The appearance of poles in the two--particle 
vertex functions $T^{(cf)}$ and $T^{(cb)}$, which signals the formation 
of collective states, may be expected to influence the behavior of 
the system in a major way.

\subsection{Self--consistent formulation: CTMA}
\begin{figure}
\vspace*{-0cm}
\centerline{\psfig{figure=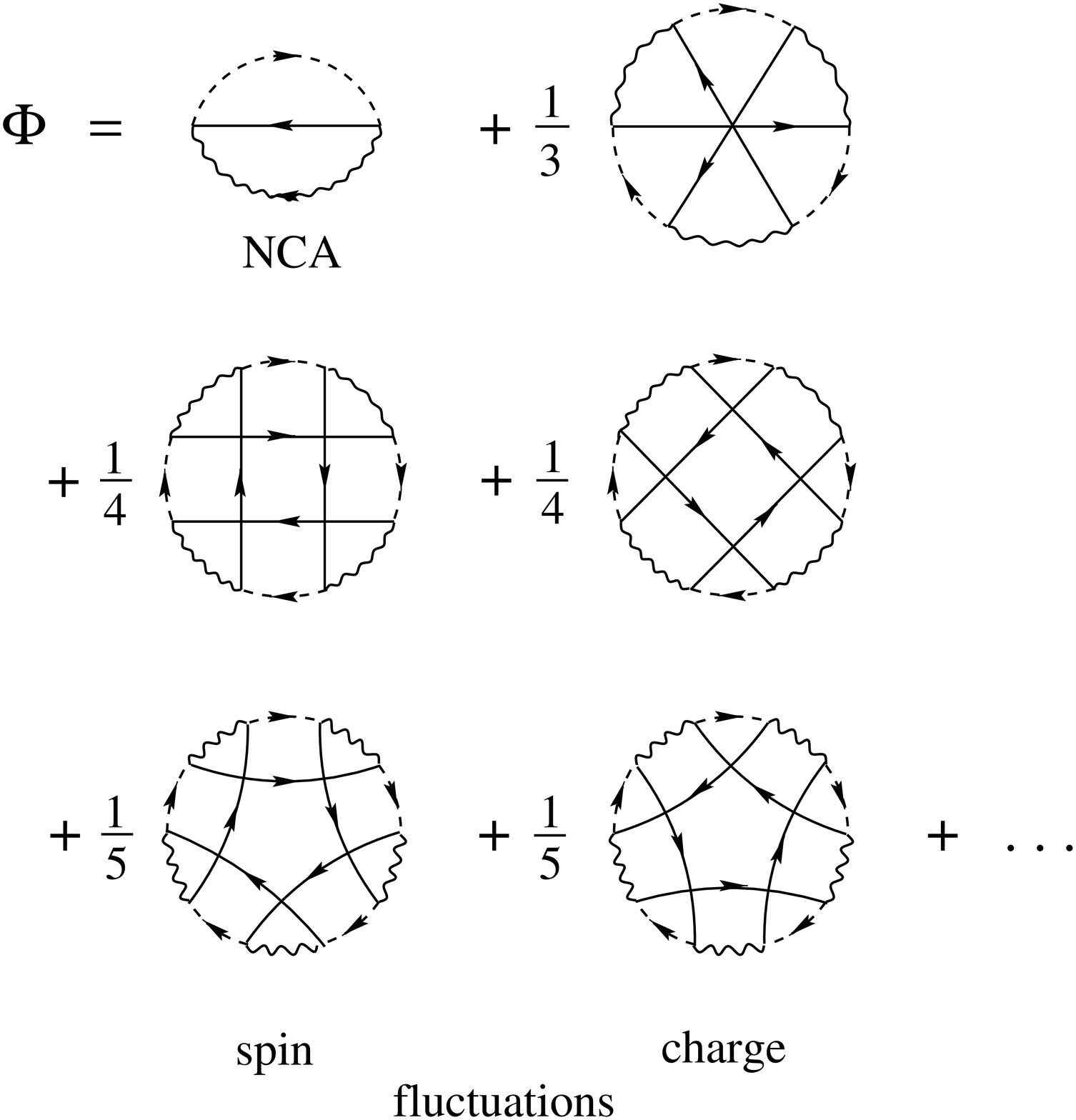,width=\linewidth}}
\vspace*{0.6cm}
\caption{ 
Diagrammatic representation of the
Luttinger--Ward functional generating the conserving
T--matrix approximation (CTMA). The terms with the conduction 
electron lines running clockwise (labeled ``spin fluctuations'') generate 
the T--matrix $T^{(cf)}$, while the terms with the conduction electron 
lines running counter--clockwise (labeled ``charge fluctuations'')
generate the T--matrix $T^{(cb)}$.
\label{CTMA}}
\end{figure}

On the level of approximation considered so far, the description is
not yet consistent: In the limit of zero temperature the spectral
weight of $T^{(cf)}$ and $T^{(cb)}$ at negative frequencies $\Omega$
should be strictly zero (threshold property). Nonvanishing 
spectral weight at $\Omega < 0$ like   
a pole contribution for negative $\Omega$ in
$T^{(cf)}$ or $T^{(cb)}$ would lead to a diverging contribution to the
self--energy, which is unphysical. However, 
recall that a minimum requirement on the approximation used is the
preservation of gauge symmetry. This requirement is not met when the
integral kernel of the $T$-matrix equation is approximated by the NCA
result. Rather, the approximation should be generated
from a Luttinger--Ward functional. The corresponding generating 
functional is shown in Fig.~\ref{CTMA}. It is defined as the 
infinite series of all vacuum skeleton diagrams which consist
of a single ring of auxiliary particle propagators, where each
conduction electron line spans at most two hybridization vertices.
As shown in appendix A by means of a cancellation theorem, the
CTMA includes, at any given loop order, all infrared singular
contributions to leading and subleading order in the frequency
$\omega$. 
The first diagram of the infinite series of CTMA terms
corresponds to NCA (Fig.~\ref{NCA}). The diagram containing two 
boson lines is excluded, since it is not a skeleton. Although the 
spirit of the present theory is different from a large $N$ 
expansion, it should be noted that the sum of the $\Phi$ diagrams 
containing up to four boson lines includes all terms of a $1/N$ 
expansion up to $O(1/N^2)$ \cite{anders.94}. By functional 
differentiation with respect to the conduction electron Green's 
function and the pseudofermion or the slave boson propagator, 
respectively, the shown $\Phi$ functional generates the ladder 
approximations $T^{(cf)}$, $T^{(cb)}$ 
for the total conduction electron--pseudofermion vertex
function and for the total conduction electron--slave boson 
vertex function (Fig.~\ref{tmat}). The auxiliary particle
self--energies are obtained in the conserving scheme as the 
functional derivatives of $\Phi$ with respect to $G_f$ or
$G_b$, respectively (Eq.~(\ref{fderiv})). This defines a set 
of self--consistency equations, which we term 
conserving T--matrix approximation (CTMA), 
where the self--energies are given as  
nonlinear and nonlocal (in time) functionals of the Green's functions,
while the Green's functions are in turn expressed in terms of the
self--energies. 
The solution of these equations requires that the T--matrices have vanishing 
spectral weight at negative COM frequencies $\Omega$. Indeed, the
numerical evaluation shows that the poles of $T^{(cf)}$ and $T^{(cb)}$ 
are shifted to $\Omega = 0$ by self--consistency, where they merge 
with the continuous spectral weight present for $\Omega >0$, thus 
renormalizing the threshold exponents of the auxiliary spectral functions.

In the following we give explicitly the self--consistent equations 
which determine the auxiliary particle self--energies within CTMA. 
For that purpose, it is useful to define conduction electron--fermion 
and conduction electron--boson vertex functions  
$T^{(cf)\,(\pm)}$, $T^{(cb)\,(\pm)}$ without ($+$) or 
with ($-$) an alternating sign between terms with even and odd number 
of rungs (compare Fig. \ref{tmat}). In the Matsubara representation,
these vertex functions, to be labelled ``even'' ($+$) and ``odd'' ($-$) 
below, are given by the following Bethe--Salpeter equations:
\widetext
\top{-2.8cm}
\begin{eqnarray}
T^{(cf)\,(\pm)\,\mu}_{\phantom{(f)\,(\pm)\,}\sigma ,\tau}
(i\omega _n, i\omega _n ', i\Omega _n) &=&
U^{(cf)\,\mu}_{\phantom{(f)\,}\sigma ,\tau}
(i\omega _n, i\omega _n ', i\Omega _n) 
\label{cftmatpm}\\
&\pm& V^2\frac{1}{\beta}\sum _{\omega _n''}
G_{b\bar\mu}(i\omega _n + i\omega _n '' - i\Omega _n ) 
G_{f\sigma}(i\omega _n'') \ G^0_{c\mu\tau}(i\Omega _n -i\omega _n '')\ 
T^{(cf)\,(\pm)\,\mu}_{\phantom{(f)\,(\pm)\,}\tau ,\sigma}
(i\omega _n '', i\omega _n ', i\Omega _n ) 
\nonumber \\
U^{(cf)\,\mu}_{\phantom{(f)\,}\sigma ,\tau}
(i\omega _n, i\omega _n ', i\Omega _n) &=&
-V^4\frac{1}{\beta}\sum _{\omega _n''}
G_{b\bar\mu}(i\omega _n + i\omega _n '' - i\Omega _n ) 
G_{f\sigma}(i\omega _n'') \ G^0_{c\mu\tau}(i\Omega _n -i\omega _n '')\ 
G_{b\bar\mu}(i\omega _n ' + i\omega _n '' - i\Omega _n )
\label{ucf}
\end{eqnarray}\noindent
and
\begin{eqnarray}
T^{(cb)\,(\pm)\sigma}_{\phantom{(b)(\pm)}\mu ,\nu}
(i\omega _n, i\omega _n ', i\Omega _n ) &=&
U^{(cb)\,\sigma}_{\phantom{(b)}\mu ,\nu}
(i\omega _n, i\omega _n ', i\Omega _n ) 
\label{cbtmatpm}\\
&\pm& V^2\frac{1}{\beta}\sum _{\omega _n''}
G_{f\sigma}(i\omega _n + i\omega _n '' - i\Omega _n ) 
G_{b\bar\mu}(i\omega _n'') \ G^0_{c\nu\sigma}(i\omega _n ''-i\Omega _n )\ 
T^{(cb)\,(\pm)\sigma}_{\phantom{(b)(\pm)}\nu ,\mu}
(i\omega _n '', i\omega _n ', i\Omega _n ) \nonumber\\
U^{(cb)\,\sigma}_{\phantom{(b)}\mu ,\nu}
(i\omega _n, i\omega _n ', i\Omega _n ) &=&
-V^4\frac{1}{\beta}\sum _{\omega _n''}
G_{f\sigma}(i\omega _n + i\omega _n '' - i\Omega _n ) 
G_{b\bar\mu}(i\omega _n'') \ G^0_{c\nu\sigma}(i\omega _n ''-i\Omega _n )\ 
G_{f\sigma}(i\omega '_n + i\omega _n '' - i\Omega _n ). 
\label{ucb}
\end{eqnarray}
\bottom{-2.7cm}
\narrowtext
\noindent
Note that, in addition to the alternating sign, 
these vertex functions differ from
the T--matrices defined in Eqs.~(\ref{cftmateq}), (\ref{cbtmateq})
in that they contain only terms with two or more rungs, since
the inhomogeneous parts $U^{(cf)}$ and $U^{(cb)}$ represent terms with
two bosonic or fermionic rungs, respectively. The terms with a 
single rung correspond to the NCA diagrams and are evaluated separately 
(see below).

The spin degrees of freedom of $T^{(cf)\,(\pm)}$ are uniquely 
determined by the spin indices $\sigma$, $\tau$ of the ingoing 
conduction electron and pseudofermion lines (Fig.~\ref{tmat}). 
It is 
\widetext
\begin{figure}
\vspace*{-0cm}
\centerline{\psfig{figure=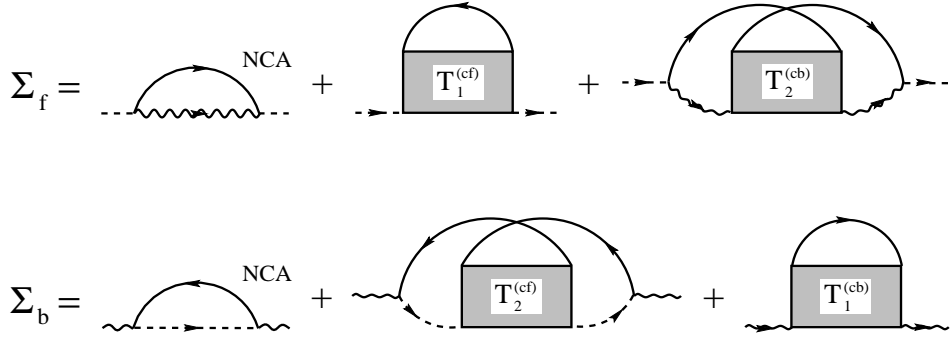,width=0.7\linewidth}}
\vspace*{0.6cm}
\caption{Diagrammatic representation of the CTMA expressions for
pseudoparticle self--energies $\Sigma _f$ and $\Sigma _b$.
The first term drawn on the righthand side
of $\Sigma _f$ and $\Sigma _b$, respectively, is the NCA diagram. 
The diagrammatic parts $T^{(cf)}_{1,2}$, $T^{(cb)}_{1,2}$ are 
explained in the text, Eqs.~(\ref{tcf1})--(\ref{tcb2}).
\label{selfenergyCTMA}}
\end{figure}
\narrowtext
instructive to note that in the spin $S=1/2$ case ($N=2$)
the singlet and triplet vertex functions  
(which correspond to the two--particle Green's functions in the singlet
channel, $\phi ^{\;s} \sim \sum _\sigma 
\langle T\{(c_{\sigma}f_{-\sigma}-c_{-\sigma}f_{\sigma})
           (c^{\dag}_{\sigma}f^{\dag}_{-\sigma}-
            c^{\dag}_{-\sigma}f^{\dag}_{\sigma})\}\rangle$,
and in the triplet channel with magnetic quantum number $m=0,\pm 1$,  
$\phi ^{\;t}_{m=0} \sim \sum _\sigma 
\langle T\{(c_{\sigma}f_{-\sigma}+c_{-\sigma}f_{\sigma})
        (c^{\dag}_{\sigma}f^{\dag}_{-\sigma}+
         c^{\dag}_{-\sigma}f^{\dag}_{\sigma})\}\rangle$,
$\phi ^{\;t}_{m=\pm 1} \sim  
\langle T\{c_{\pm\frac{1}{2}}f_{\pm\frac{1}{2}}
           c^{\dag}_{\pm\frac{1}{2}}f^{\dag}_{\pm\frac{1}{2}}\}\rangle$, 
respectively)
may be identified in the following way,
\begin{eqnarray}
T^{(cf)\, s} &=&        \sum _{\sigma } T^{(cf)\, (-)}_{\sigma , -\sigma} 
\\
T^{(cf)\, t}_{m=0} &=&    \sum _{\sigma } T^{(cf)\, (+)}_{\sigma , -
\sigma} \\
T^{(cf)\, t}_{m=\pm 1}&=& \phantom{\sum _{\sigma }} 
                        T^{(cf)\, (+)}_{\pm\frac{1}{2},\pm\frac{1}{2} }. 
\end{eqnarray}
Analogous relations hold for the conduction electron--boson
vertex function in terms of the channel degrees of freedom $\mu$, $\nu$.
The total CTMA pseudoparticle self--energies, as derived by functional 
differentiation from the generating functional $\Phi$, Fig.~\ref{CTMA},
are shown in Fig.~\ref{selfenergyCTMA} and consist of three terms each,
\begin{eqnarray}
\Sigma _{f\sigma}(i\omega  _n) &=&\Sigma _{f\sigma}^{(NCA)}(i\omega _n )+
                             \Sigma _{f\sigma}^{(cf)} (i\omega _n ) +
                             \Sigma _{f\sigma}^{(cb)} (i\omega _n )
 \nonumber \\ \\
\Sigma _{b\bar\mu}(i\omega _n)&=&\Sigma _{b\bar\mu}^{(NCA)}(i\omega _n )+
                             \Sigma _{b\bar\mu}^{(cf)} (i\omega _n ) +
                             \Sigma _{b\bar\mu}^{(cb)} (i\omega _n )\; .
 \nonumber \\
\end{eqnarray}
The first term of $\Sigma _f$ and $\Sigma _b$ represents  
the NCA self--energies, Eqs.~(\ref{sigfNCA}), 
(\ref{sigbNCA}). The second and third terms arise from the spin and 
the charge fluctuations, respectively, and are given for pseudofermions by
\widetext
\top{-2.8cm}
\begin{eqnarray}
\Sigma _{f\sigma}^{(cf)}(i\omega _n) &=& M\;\frac{1}{\beta}\sum _{\Omega _n}
G^0_{c}(i\Omega _n-i\omega _n)
T^{(cf)}_1 (i\omega _n,i\omega _n,i\Omega _n)\\
\Sigma _{f\sigma}^{(cb)}(i\omega _n) &=& 
-M\; V^2\frac{1}{\beta ^2}\sum _{\omega _n '\omega _n''}
G^0_{c}(i\omega _n-i\omega _n')
G _{b}(i\omega _n')
T^{(cb)}_2 
(i\omega _n ',i\omega _n '',i\omega _n'+i\omega _n''-i\omega _n)
G^0_{c}(i\omega _n-i\omega _n'') G _{b}(i\omega _n'')
\end{eqnarray}
and for slave bosons by
\begin{eqnarray}
\Sigma _{b\bar\mu}^{(cf)}(i\omega _n ) &=& 
-N\; V^2\frac{1}{\beta ^2}\sum _{\omega _n '\omega _n''}
G^0_{c}(i\omega _n'-i\omega _n)
G _{f}(i\omega _n')
T^{(cf)}_2 (i\omega _n ',i\omega _n '',i\omega _n'+i\omega _n''-i\omega _n)
G^0_{c}(i\omega _n''-i\omega _n)
G _{f}(i\omega _n'')\\
\Sigma _{b\bar\mu}^{(cb)}(i\omega _n) &=& N\;\frac{1}{\beta}\sum _{\Omega _n}
G^0_{c}(i\omega _n -i\Omega _n) T^{(cb)}_1 
(i\omega _n,i\omega _n,i\Omega _n)\; ,
\end{eqnarray}
where the vertex functions appearing in these expressions are defined as
\begin{eqnarray}
T^{(cf)}_1 &=& \frac{N+1}{2}\;T^{(cf)\,(+)} + \frac{N-1}{2}\; T^{(cf)\,(-)}
               -N\, U^{(cf)}\label{tcf1}\\
T^{(cf)}_2 &=& \frac{N+1}{2}\;T^{(cf)\,(+)} - \frac{N-1}{2}\; T^{(cf)\,(-)}
               - U^{(cf)} \label{tcf2}
\end{eqnarray}
\begin{eqnarray}
T^{(cb)}_1 &=& \frac{M-1}{2}\; T^{(cb)\,(+)} + \frac{M+1}{2}\; 
               T^{(cb)\,(-)}-M\, U^{(cb)}\label{tcb1}\\
T^{(cb)}_2 &=& \frac{M-1}{2}\; T^{(cb)\,(+)} - \frac{M+1}{2}\; 
               T^{(cb)\,(-)}-U^{(cb)}\; .\label{tcb2}
\end{eqnarray}
\bottom{-2.7cm}
\narrowtext
These combinations of the even and odd vertex
functions ensure the proper spin and channel summations in the
self--energies. For the sake of clarity, the spin and channel 
indices as well as the frequency variables are not shown explicitly. 
In Eqs. (\ref{tcf1}), (\ref{tcb1}) 
the terms with two rungs, $N\, U^{(cf)}$, $M\, U^{(cb)}$, have been
subtracted, since they would generate non--skeleton self--energy diagrams.
Likewise, in Eqs. (\ref{tcf2}), (\ref{tcb2}) the two--rung terms have
been subtracted in order to avoid a double counting of terms in the
self--energies. 

We now turn to the analytic continuation to real frequencies of the
expressions derived above. 
Transforming the Matsubara summations into contour integrals shows
that integrations along branch cuts of
auxiliary particle Green's functions carry an additional factor
${\rm exp}(-\beta\lambda)$ as compared to integrations along branch 
cuts of physical Green's functions, which vanishes upon projection 
onto the the physical Fock space, $\lambda \rightarrow \infty$.
Thus, as a general rule, only integrations along branch cuts of the
$c$--electron propagators contribute to the auxiliary particle 
self--energies. Therefore, by performing the analytic continuation, 
$i\omega_n \rightarrow \omega -i0 \equiv \omega$ in all frequency
variables, we obtain the advanced pseudofermion self--energy,
\widetext
\top{-2.8cm}
\begin{eqnarray}
\Sigma _{f\sigma}^{(cf)}(\omega ) &=& M
\int \frac{ d\varepsilon}{\pi}\;f(\varepsilon -\omega)\;
A^0_{c}(\varepsilon -\omega)\; \pi{\cal N}(0)
T^{(cf)}_1 (\omega ,\omega ,\varepsilon ) \label{sigfcontinued} \\
\Sigma _{f\sigma}^{(cb)}(\omega ) &=& 
-M\; \Gamma \int \frac{ d\varepsilon}{\pi}\int \frac{ d\varepsilon '}{\pi}
\;f(\varepsilon -\omega)\;f(\varepsilon ' -\omega)
A^0_{c}(\omega -\varepsilon )
G _{b}(\varepsilon )\; 
\pi{\cal N}(0) T^{(cb)}_2 (\varepsilon ,\varepsilon ' ,
\varepsilon +\varepsilon ' -\omega )
A^0_{c}(\omega -\varepsilon ') G _{b}(\varepsilon ' )
\end{eqnarray}
and the advanced slave boson self--energy,
\begin{eqnarray}
\Sigma _{b\bar\mu}^{(cf)}(\omega ) &=& 
-N\; \Gamma \int \frac{ d\varepsilon}{\pi}\int \frac{ d\varepsilon '}{\pi}
\;f(\varepsilon -\omega)\;f(\varepsilon '-\omega )
A^0_{c}(\varepsilon -\omega )
G _{f}(\varepsilon )\;
\pi{\cal N}(0)T^{(cf)}_2 (\varepsilon ,\varepsilon ' ,
\varepsilon +\varepsilon ' -\omega )
A^0_{c}(\varepsilon ' -\omega)
G _{f}(\varepsilon ' ) \\
\Sigma _{b\bar\mu}^{(cb)}(\omega ) &=& -N
\int \frac{ d\varepsilon}{\pi}\;f(\varepsilon -\omega)\;
A^0_{c}(\omega -\varepsilon )\;
\pi{\cal N}(0) T^{(cb)}_1 (\omega ,\omega ,\varepsilon ) ,
\label{sigbcontinued}
\end{eqnarray}
where the vertex functions are given by Eqs. (\ref{tcf1}), (\ref{tcf2}) and 
(\ref{tcb1}), (\ref{tcb2}) with
\begin{eqnarray}
T^{(cf)\,(\pm)\,\mu}_{\phantom{(f)\,(\pm)\,}\sigma ,\tau}
(\omega , \omega ' , \Omega ) &=&
U^{(cf)\,(\pm)\,\mu}_{\phantom{(f)\,(\pm)\,}\sigma ,\tau}
(\omega , \omega ' , \Omega )
\nonumber\\
&\pm& (-\Gamma) \int \frac{ d\varepsilon}{\pi}\;f(\varepsilon -
\Omega)  
G_{b\bar\mu}(\omega+\varepsilon-\Omega)  
G_{f\sigma}(\varepsilon )  A^0_{c\mu\tau}(\Omega -\varepsilon)  
T^{(cf)\,(\pm)\,\mu}_{\phantom{(f)\,(\pm)\,}\tau ,\sigma}
(\varepsilon , \omega ', \Omega  ) 
\label{tcfcontinued}\\
\pi{\cal N}(0)\; U^{(cf)\,(\pm)\,\mu}_{\phantom{(f)\,(\pm)\,}\sigma ,\tau}
(\omega , \omega ' , \Omega ) &=&
+\Gamma ^2 \int \frac{ d\varepsilon}{\pi}\;f(\varepsilon -\Omega) 
G_{b\bar\mu}(\omega +\varepsilon -\Omega) 
G_{f\sigma}(\varepsilon )  A^0_{c\mu\tau}(\Omega -\varepsilon) 
G_{b\bar\mu}(\omega '  +\varepsilon -\Omega)  
\end{eqnarray}
and
\begin{eqnarray}
T^{(cb)\,(\pm)\sigma}_{\phantom{(b)(\pm)}\mu ,\nu}
(\omega , \omega ', \Omega ) &=&
U^{(cb)\,(\pm)\sigma}_{\phantom{(b)(\pm)}\mu ,\nu}
(\omega , \omega ', \Omega )
\nonumber\\
&\pm& (+\Gamma) \int \frac{ d\varepsilon}{\pi}\;f(\varepsilon -\Omega )
G_{f\sigma}(\omega +\varepsilon -\Omega ) 
G_{b\bar\mu} (\varepsilon )
A^0_{c\nu\sigma}(\varepsilon -\Omega )\ 
T^{(cb)\,(\pm)\sigma}_{\phantom{(b)(\pm)}\nu ,\mu}
(\varepsilon , \omega ', \Omega ) \\
\pi{\cal N}(0)\; U^{(cb)\,(\pm)\sigma}_{\phantom{(b)(\pm)}\mu ,\nu}
(\omega , \omega ', \Omega ) &=&
-\Gamma ^2\int \frac{ d\varepsilon}{\pi}\;f(\varepsilon -\Omega) 
G_{f\sigma}(\omega + \varepsilon -\Omega ) 
G_{b\bar\mu}(\varepsilon ) A^0_{c\nu\sigma}(\varepsilon -\Omega)\ 
G_{f\sigma}(\omega '+ \varepsilon -\Omega ) \label{tcbcontinued} 
\end{eqnarray}
\bottom{-2.7cm}
\narrowtext
\noindent
In the above expressions, like in the NCA equations 
(\ref{sigfNCA})--(\ref{gdNCA}),
we have used the dimensionless conduction electron spectral density, 
$A_{c}^0(\omega )=\frac{1}{\pi}\, {\rm Im}G_{c\mu\sigma}^0(\omega -i0)/
{\cal N}(0)$, and we have suppressed obvious spin and channel indices. 
All frequency variables are to be understood as the limit
$\omega\equiv \omega -i0$. \\
The equations (\ref{sigfcontinued})--(\ref{sigbcontinued}), 
supplemented by the vertex functions Eqs. (\ref{tcf1})--(\ref{tcb2}), 
(\ref{tcfcontinued})--(\ref{tcbcontinued}) form, together with the 
NCA contributions Eqs. (\ref{sigfNCA}), (\ref{sigbNCA}) and the 
definitions of the auxiliary particle Green's functions, 
Eqs.~(\ref{green}), (\ref{green0}), the closed set of self--consistent
CTMA equations \cite{kroha.97}. 
It is seen that in these equations only those branches of the 
T--matrix vertex functions appear which are advanced with respect to all 
three frequency variables, although in general the T--matrix consists of
$2^3$ independent analytical branches. This simplification is a 
conse-
\widetext 
\begin{figure}
\hspace*{0cm}\psfig{figure=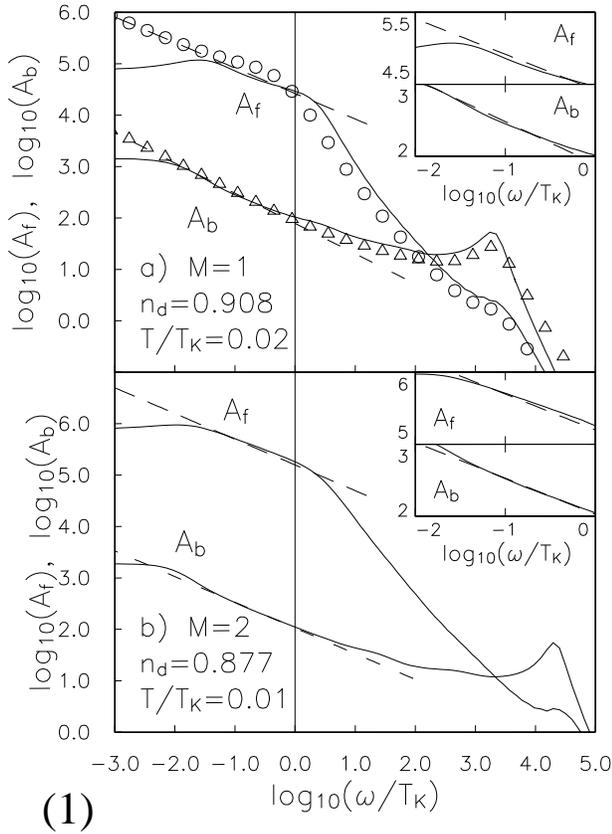,width=0.45\linewidth}

\vspace*{-11cm}
\hfill{\psfig{figure=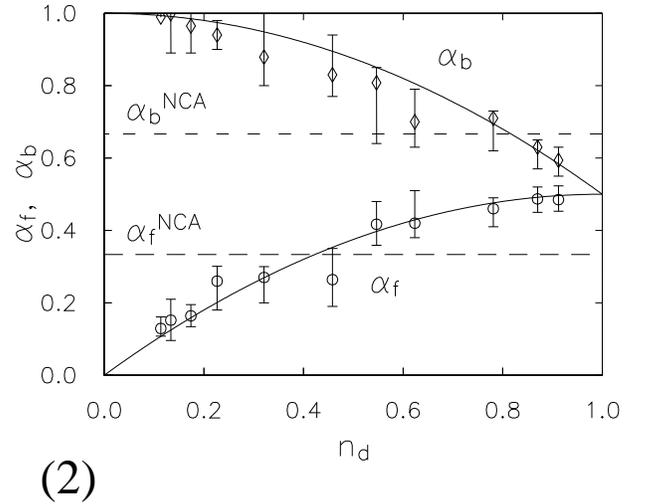,width=0.45\linewidth}}
\vspace*{0.6cm}
\caption{ 
(1) Pseudofermion and slave boson spectral functions $A_f$ and $A_b$
in the Kondo regime ($N=2$; 
$E_d=-0.05$, $\Gamma =0.01$ in units of the half--bandwidth $D$), 
for a) the single--channel ($M=1$) and b) the
multi--channel ($M=2$) case.
In a) the symbols represent the results of NRG for the same 
parameter set, $T=0$. The slopes of the dashed lines indicate the exact
threshold exponents as derived in section III B for $M=1$ and as given by 
conformal field theory for $M=2$. The insets show magnified power law 
regions. \ 
(2) CTMA results (symbols with error bars) for the threshold 
exponents $\alpha _f$ and $\alpha _b$ of $A_f$ and $A_b$, ($N=2$, 
$M=1$). Solid lines: exact values (section III B), dashed lines: NCA results 
(section V B).
\label{spectralfb}}
\end{figure}
\narrowtext
quence of the exact projection onto the physical sector of Fock space. 
Inspection of the analytically continued CTMA equations also shows that 
the slave boson self--energy is obtained from the pseudofermion 
self--energy, including the proper signs, by simply replacing 
$G_f \leftrightarrow G_b$ and inverting the frequency argument of 
$A_c^0$ in all expressions.

Eqs. (\ref{sigfcontinued})--(\ref{tcbcontinued}) 
may be rewritten in terms of the spectral functions without
threshold, $\tilde A_{f,b}$, in a straight--forward way 
as explained in section V C, thus avoiding divergent statistical 
factors in the $d$--electron Green's function.

The self--consistent solutions are obtained by first solving the
linear Bethe--Salpeter equations for the T--matrices
by matrix inversion, computing the auxiliary particle 
self--energies from $T^{(cf)}$ and $T^{(cb)}$, and then constructing
the fermion and boson Green's functions from the respective 
self--energies. This process is iterated until convergence is reached.  
We have obtained reliable solutions  
down to temperatures of the order of at least $10^{-2} T_K$
both for the single-channel and for the two-channel Anderson
model. Note that $T_K\rightarrow 0$ in the Kondo limit; 
in the mixed valence and empty impurity regimes, significantly lower
temperatures may be reached, compared to the low temperature scale 
of the model.

\subsection{Results for the auxiliary particle spectral functions}
As shown in Fig.~\ref{spectralfb} (1) a), 
the auxiliary particle spectral functions obtained from CTMA 
\cite{kroha.97} are in 
good agreement with the results of a numerical renormalization 
group (NRG) calculation \cite{costi.94}
(zero temperature results), given the uncertainties in the 
NRG at higher frequencies. Typical behavior in the Kondo regime 
is obtained: a broad peak of width $\sim \Gamma$ 
in $A_b$ at $\omega\simeq |E_d|$, 
representing the hybridizing $d$--level and a structure 
in $A_f$ at $\omega \simeq T_K$. Both functions display power law 
behavior at frequencies
below $T_K$, which at finite $T$ is cut off at the scale $\omega
\simeq T$. The exponents extracted from the frequency range
$T<\omega<T_K$ of our finite $T$ results  
compare well with the exact result also shown
(see insets of Fig.~\ref{spectralfb} (1) $a$)). A similar analysis has been 
performed for a number of parameter sets spanning the complete range of
$d$--level occupation numbers $n_d$. The extracted power law exponents
are shown in Fig.~\ref{spectralfb} (2), 
together with error bars estimated 
from the finite frequency ranges over which the fit was made.
The comparatively large error bars
in the mixed valence regime arise because here spin flip and
charge fluctuation processes, described by the poles in $T^{(cf)}$ and
$T^{(cb)}$, respectively, are of equal importance, impeding the 
convergence of the numerical procedure. In this light, the agreement
with the exact results (solid curves) is very good, 
the exact value lying within the error bars or very close 
in each case.
\begin{figure}
\hspace*{0.03\linewidth}\psfig{figure=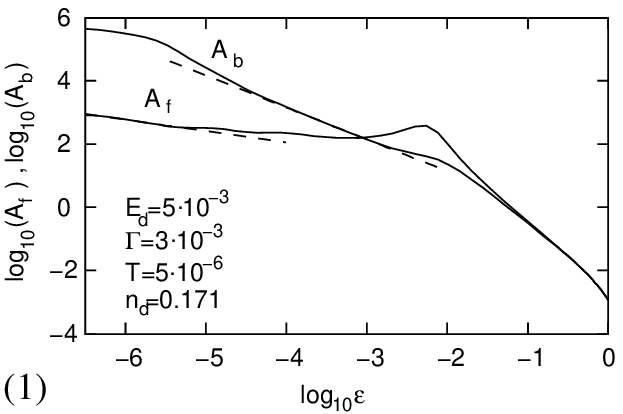,width=0.42\textwidth}

\hspace*{0.04\linewidth}{\psfig{figure=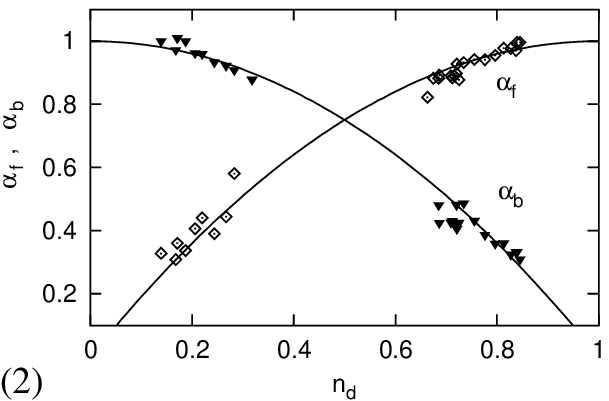,width=0.42\textwidth}}

\vspace*{0.6cm}
\caption{ 
(1) Pseudofermion and slave boson spectral functions $A_f$ and $A_b$
for the X-ray problem ($N=1$, $M=1$; 
$E_d=0.005$, $\Gamma =0.003$ in units of the half--bandwidth $D$)
[48]. The slopes of the dashed lines indicate the exact
threshold exponents as derived in section III B.  \ 
(2) CTMA results (symbols) for the threshold 
exponents $\alpha _f$ and $\alpha _b$ of the spectral functions
$A_f$ and $A_b$, ($N=1$, $M=1$). 
Solid lines: exact values (section III B).
\label{spectralfbN1}}
\end{figure}

In the multi--channel case ($N\geq 2$, $M\geq N$) NCA has been
shown \cite{coxruck.93} to reproduce asymptotically the correct 
threshold exponents, $\alpha _f = M/(M+N)$, $\alpha _b = N/(M+N)$,
in the Kondo limit. Calculating the T--matrices using NCA Green's 
functions (as discussed in the single--channel case) we find again
a pole in the singlet channel of $T^{(cf)}$. However, in this case 
the CTMA does not renormalize the NCA exponents 
in the Kondo limit of the two--channel model, i.e.~the threshold 
exponents obtained from the CTMA solutions are very close to the
exact ones, $\alpha _f =1/2$, $\alpha _b=1/2$, as shown in 
Fig.~\ref{spectralfb} (1) b). 

We note in passing that we have also solved the spinless case ($N=1$, $M=1$)
using the auxiliary particle method \cite{schauerte.00}. The Anderson impurity
model, Eqs.~(\ref{2.5})--(\ref{2.6}), then
reduces to a noninteracting resonant level model.
In this sense, the spinless case may be seen as the most extreme case 
of a quantum impurity model with a Fermi liquid ground state. It does, 
however, remain non--trivial in the slave particle representation 
Eq.~(\ref{sbhamilton}) and, therefore, constitutes a test case for
the description of Fermi liquid behavior within the auxiliary particle
method. As discussed in section III B, the auxiliary particle 
spectral functions $A_f$, $A_b$ in general display an infrared threshold 
with power law behavior which is induced by an orthogonality catastrophy
analogous to that of the X-ray problem. In the spinless case, $A_f$ and $A_b$
have been shown to correspond precisely to the photoemission and 
X-ray absorption spectral densities of the X-ray problem, 
respectively \cite{schauerte.00}.
The results of the CTMA for the $N=1$, $M=1$ case are
shown in Fig.~\ref{spectralfbN1}. Good quantitative agreement of the
X-ray exponents extracted from the numerical evaluation of the 
CTMA equations is found in the regions $n_d \stackrel {<}{\sim} 1$
and $n_d \stackrel {>}{\sim} 0$
which may be identified with the regions of a strong (scattering phase
shift $\eta _{\sigma} \simeq \pi /2$) and a weak potential
(scattering phase shift $\eta _{\sigma} \simeq 0$) scatterer.

The agreement of the CTMA exponents with their exact values
in the Kondo, mixed valence and empty impurity regimes of the Spin 1/2
single--channel model, in the spinless model and in the Kondo regime of the
two--channel model may be taken as
evidence that the T--matrix approximation correctly 
describes both the Fermi liquid ($N=1,2$, $M=1$) and the 
non--Fermi liquid ($N=2$, $M=1,2$)
regimes of the SU(N)$\times$SU(M) Anderson model.
Therefore, we expect the CTMA to correctly describe  
physically observable quantities of the SU(N)$\times$SU(M) Anderson 
impurity model as well.

\subsection{Results for physical quantities: Spin susceptibility}
\begin{figure}
\vspace*{-0cm}
\centerline{\psfig{figure=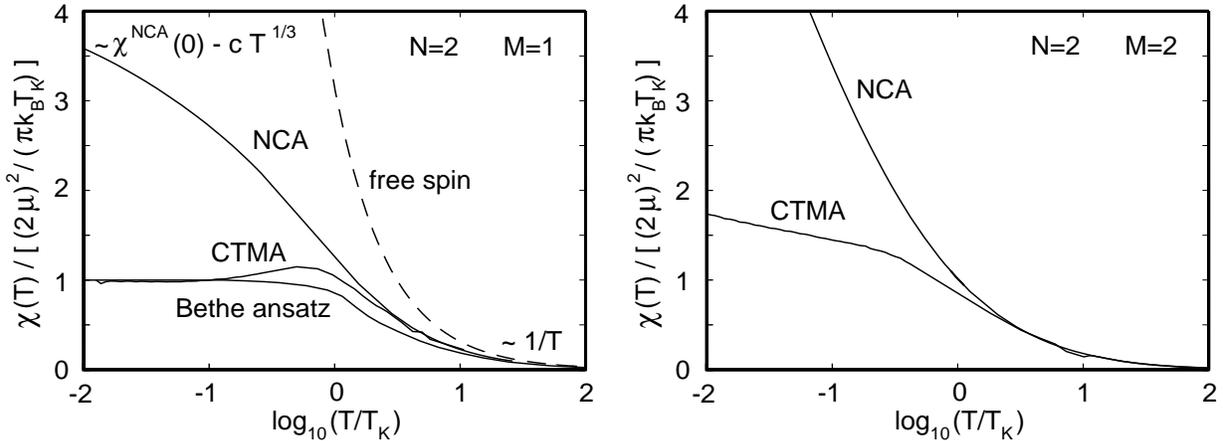,width=0.9\linewidth}}
\vspace*{0.6cm}
\caption{
Static susceptibility of the single-channel ($N=2$, $M=1$) and the 
two-channel ($N=2$, $M=2$) Anderson impurity model in the Kondo regime 
($E_d=-0.8D$, $\Gamma = 0.1D$, Land\'e factor $g=2$). In the single-channel
case, the CTMA and NCA results 
are compared with the Bethe ansatz result for the Kondo model [14],
where the CTMA as well as the Bethe ansatz curves are scaled by 
their respective Kondo temperatures [14]. 
The CTMA susceptibility is universal in that it depends 
only on the ratio $T/T_K$ for various values of the parameters of the
Anderson model (not shown).
\label{susc}}
\end{figure}

We have calculated
the static spin susceptibility $\chi$ of the Anderson
model in the Kondo regime by solving the CTMA equations 
in a finite magnetic field $H$ coupled to the impurity spin and taking 
the derivative of the magnetization 
$M = \frac{1}{2}g \mu _B \langle n_{f\uparrow} - n_{f\downarrow}\rangle$ 
with respect to $H$. 
The resulting $\chi (T) = (\partial M / \partial H )_T$ is shown 
in Fig.~\ref{susc} both for the single--channel case ($N=2$,
$M=1$) and for the two--channel Anderson Model ($N=2$, $M=2$).
It is seen that in the single-channel case 
CTMA correctly reproduces the constant Pauli susceptibility 
(Fermi liquid behavior) below $T_K$, while NCA gives an incorrect,
nonanalytic temperature dependence of $\chi(T)-\chi(0) \propto - T^{1/3}$
at low $T$. 
In the two-channel case CTMA describes the  
non-Fermi liquid behavior, i.e. the 
the exact \cite{Wieg83} logarithmic temperature dependence of the 
susceptibility below the Kondo scale $T_K$. In contrast, the NCA solution 
recovers the logarithmic behavior only far below $T_K$. Other physical 
quantities have been calculated for the Anderson model at low temperatures 
as well and will be published in forthcoming work \cite{kirchner.01}.

\widetext
\narrowtext

\section{Anderson model at finite U: Generalized NCA and CTMA}
\label{anderson model}

The single channel Anderson model at finite U may be represented in
terms of the pseudofermion operators $f_\sigma$ and slave boson
operators $a,b$ (discussed in the beginning of section III) as
\begin{eqnarray}
H &=& 
H_c+E_d(\Sigma_\sigma f_\sigma^+f_\sigma + 2a^+a)  + Ua^+a \nonumber \\
  &+& V\Sigma_{\vec
k\sigma}\Big[c_{\vec k\sigma}^+(b^+f_\sigma + \eta_\sigma 
f_{-\sigma}^+a)+h.c.\Big]
\label{7.1}
\end{eqnarray}
subject to the constraint
\begin{equation}
\sum_\sigma f_\sigma^+f_\sigma + a^+a + b^+b = 1
\end{equation}
There are now two bosons, a ``light'' boson b (as in the case
$U\rightarrow\infty$ considered before) and a ``heavy'' boson $a$.
As seen from Eq.~(\ref{7.1}), the strong two--particle interaction $U$ is 
transformed into a potential term by this representation. 
The corresponding diagrams are made of propagator lines for the light
boson (wiggly line, as before) and for the heavy boson (zig-zag line) as
well as a new vertex, where incoming conduction electron lines and
pseudofermion lines merge into an outgoing heavy boson line.

\subsection{Generating functional}
\label{generating}

\begin{figure}
\vspace*{-0cm}
\centerline{\psfig{figure=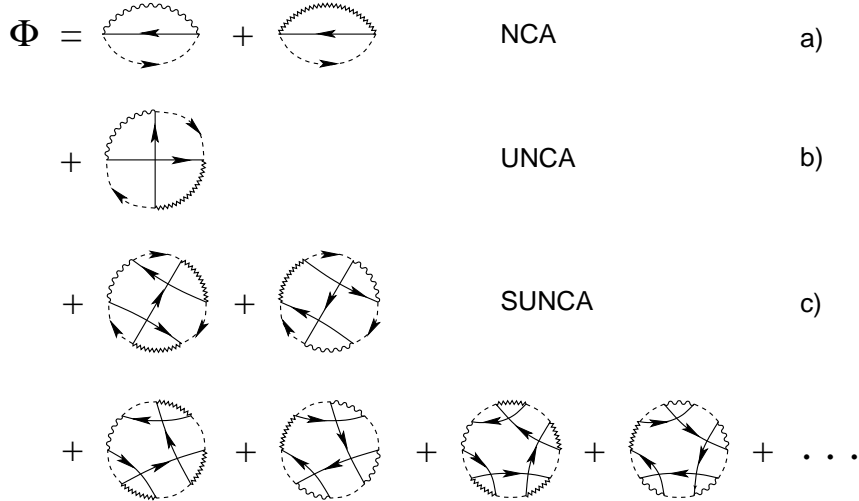,width=0.63\linewidth}}
\vspace*{0.6cm}
\caption{Diagrammatic representation of the generating functional for the 
Anderson impurity model at finite $U$. Solid, dashed, wiggly and zig-zag
lines correspond to conduction electron $c$, pseudofermion $f$, 
``light'' boson $b$, and ``heavy'' boson $a$ propagators.
a) NCA including light
(empty impurity) and heavy (doubly occupied impurity) boson lines.
b) UNCA containing, in addition to an infinite number of
(bare) hybridization processes into the empty impurity state 
(non-skeleton diagrams), a single hybridization into the doubly occupied 
impurity state. 
c) Symmetrized, finite-$U$ NCA (SUNCA): The infinite series of all diagrams
of a)--c) generates an approximation where the hybridization processes
into the empty and doubly occupied state are treated in a symmetric
way, i.e. for each contribution with a bare light boson line there is a
corresponding contribution with the bare light boson line replaced by a
bare heavy boson line, and vice versa. 
\label{SUNCA}}
\end{figure}

It seems straightforward to define a generalization of NCA for finite $U$
by adding to the second order skeleton diagram for the generating functional
$\Phi$ a second one where the light boson is replaced by the heavy boson
(see Fig. \ref{SUNCA} a)). This approximation has been considered sometime ago
\cite{pruschke89,holmschoenh89},
where it was found to fail badly: Not even the Kondo energy scale is
recovered in the so defined approximation.  The reason for this
failure is obvious: In the Kondo regime $(n_d \sim 1)$ the local spin
is coupled to the conduction electron  spin density at the impurity
through the antiferromagnetic exchange coupling $J = V^2 (-
\frac{1}{E_d} + \frac{1}{E_d + U})$.  The two terms on the r.h.s. of
this relation arise from virtual transitions into the empty and doubly
occupied local level, which contribute equally in the symmetric case
$\mid E_d\mid = E_d+U$.  The symmetric occurrence of these two virtual
processes in all intermediate states is not included in the simple
extension of NCA proposed above.  The self-energy insertions in each
of the two diagrams contain always only one of the processes, leading
to an effective $J$ which is only one half of the correct value.  
Correspondingly, the Kondo temperature $T_K \sim \exp(-1/{\cal
N}(0)J)$ comes out to scale as the square of the correct value, which
is orders of magnitude too small.
\widetext
\narrowtext

\begin{figure}
\vspace*{-0cm}
\centerline{\psfig{figure=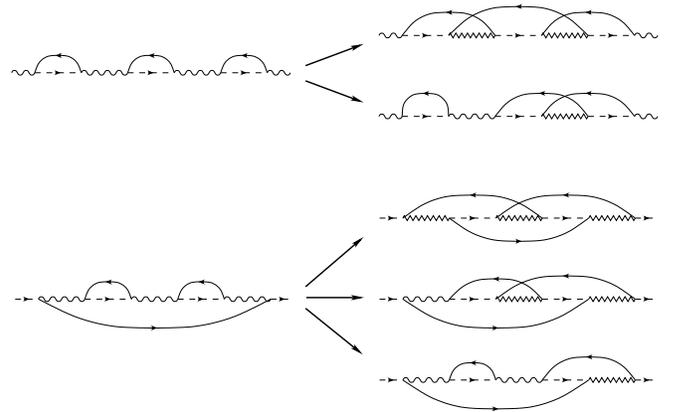,width=\linewidth}}
\vspace*{0.6cm}
\caption{Two examples of how diagrams involving hybridization into the
doubly occupied impurity state are generated from the  bare 
noncrossing diagrams of the infinite $U$ case
by replacing light with heavy boson lines (see text).
\label{example}}
\end{figure}

To correct this deficiency it is necessary to include additional
diagrams, restoring the symmetry between the two virtual processes,
excitation into the empty or into the doubly occupied impurity state.
Due to the exact projection onto the physical Hilbert space,
each diagram contributing to $\Phi$ contains exactly one auxiliary
particle ring (see, e.g. Fig.~\ref{CTMA}).
As a first step one may add the next order diagram to $\Phi$ (see
Fig.~\ref{SUNCA} b)).  
As will be seen below, this approximation, later referred
to as UNCA, helps to recover a large part of the correct behavior of
$T_K$.  A completely symmetric treatment, however, requires the
summation of infinite classes of diagrams.  These diagrams are
generated by replacing a light boson line with a heavy boson line in
each of the bare (non-skeleton) diagrams of the elementary NCA
diagrams of Fig.~\ref{example}.  Each replacement leads to a crossing of
conduction electron lines spanning one fermion and at most two boson
lines.  As a result, the class of diagrams of $\Phi$ shown in
Fig.~\ref{SUNCA} c) are obtained.  These diagrams look similar to the CTMA
diagrams shown in Fig.~\ref{CTMA}, but contain one light boson line and an
arbitrary number of heavy boson lines, or vice versa.  Diagrams with,
for example, two light boson lines and an arbitrary number of heavy
boson lines (and conduction electron lines spanning at most one fermion
line) are reducible and do not appear.  We will call the approximation
defined by the generating functional given by the sum of the diagrams
of Figs.~\ref{SUNCA} a)--c) 
``symmetrized, finite--$U$ non--crossing approximation'' (SUNCA)
\cite{haule00,haule01}. It is
interesting to note that the symmetric treatment of light and heavy
bosons leads to diagrams for $\Phi$ of a structure similar to the ones
defining the CTMA (Fig.~\ref{CTMA}).  The above approximation
corresponding to the CTMA at $U \rightarrow \infty$, termed 
``symmetrized, finite--$U$
Conserving T-matrix Approximation (SUCTMA) is thus defined in
a natural way by summing up all skeleton diagrams with an arbitrary
number of light or heavy boson lines, dressed by conduction electron
lines spanning only one fermion line. The rational for keeping only
these crossings is that (as explained in the appendix), 
\noindent
diagrams 
containing conduction electron lines spanning two or more
fermion lines may be grouped into sets where the most singular
contributions cancel. Thus, the SUCTMA is defined by adding to the
diagrams of the SUNCA the CTMA diagrams (Fig.~\ref{CTMA}) with (only) light
boson lines or (only) heavy boson lines.  The SUCTMA equations have not
yet been evaluated.
\widetext
\begin{figure}
\vspace*{-0cm}
\centerline{\psfig{figure=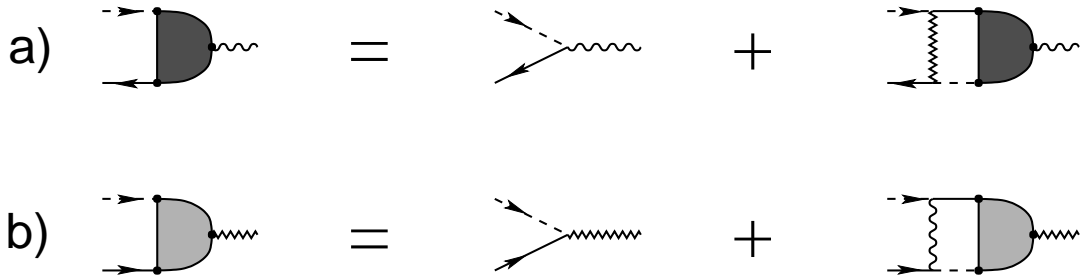,width=0.8\linewidth}}
\vspace*{0.6cm}
\caption{
Diagrammatic representation of the Bethe--Salpeter equations for a) the
renormalized light boson (empty impurity) and  b) the
renormalized heavy boson (doubly occupied impurity) vertex, as
generated by the SUNCA Luttinger-Ward functional (Fig.~11).
\label{Uvertex}}
\end{figure}
\narrowtext
\subsection{Results of SUNCA} 

\begin{figure}
\vspace*{-0cm}
\centerline{\psfig{figure=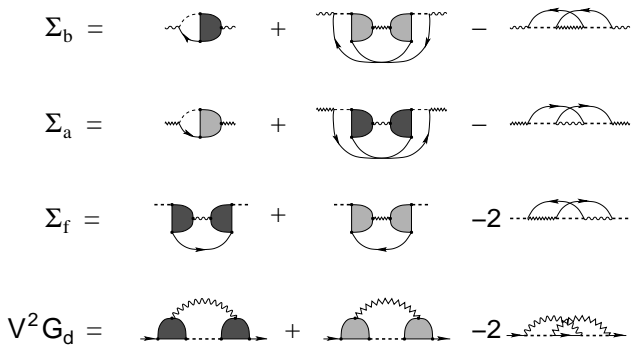,width=0.96\linewidth}}
\vspace*{0.6cm}
\caption{
Diagrammatic representation of the auxiliary particle selfenergies
of SUNCA in terms of the renormalized hybridization vertices, defined
in Fig.~12. Note that in each line the third diagram is subtracted
in order to avoid double counting of terms within the first two diagrams.  
\label{Uselfenergy}}
\end{figure}

The pseudoparticle self-energies $\Sigma_a, \Sigma_b$ and $\Sigma_f$
as well as the local conduction electron self-energy $\Sigma_c$ are
obtained as funcional derivatives with respect to the corresponding
Green's functions.  They can be expressed in terms of the two vertex
functions defined in Fig.~\ref{Uvertex} as solutions of Bethe-Salpeter
equations.  It is seen that the vertex functions are obtained by
ladder summations involving a heavy boson or a light boson line as the
rung of the ladder. These three-point vertex functions depend on two
frequencies and are much less difficult to calculate than the
four-point T-matrices of the CTMA.  The full expressions of the
self-energies are given in Fig.\ref{Uselfenergy}.
The local d-electron Green's function after projection is again
proportional to $\Sigma_c$: 
\begin{equation} G_d(\omega) = \frac{1}{V^2} \Sigma_c(\omega) 
\label{eq:G_d}
\end{equation}

\begin{figure}
\centerline{\psfig{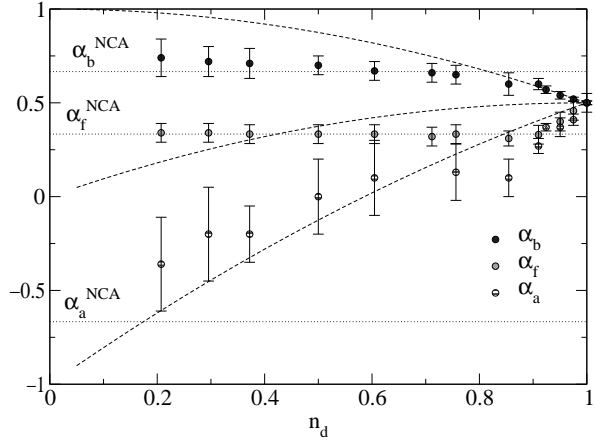}}
\vspace*{0.6cm}
\caption{Infrared threshold exponents of the auxiliary particle 
spectral functions for the case of finite $U$
in dependence of the the impurity occupation $n_d$.
Dashed curved lines: exact results (Eqs.~(96)-(98)); horizontal lines:
NCA results; data points with error bars: SUNCA results.
In the Kondo limit ($n_d \to 1$)
the exact exponents are recovered, while in the mixed valence and
empty impurity regime the SUNCA results for $\alpha _f$ and $\alpha _b$
cross over to the NCA values.
\label{exponentsU}}
\end{figure}

\begin{figure}
\vspace*{-0cm}
\hspace*{0.02\linewidth}\psfig{figure=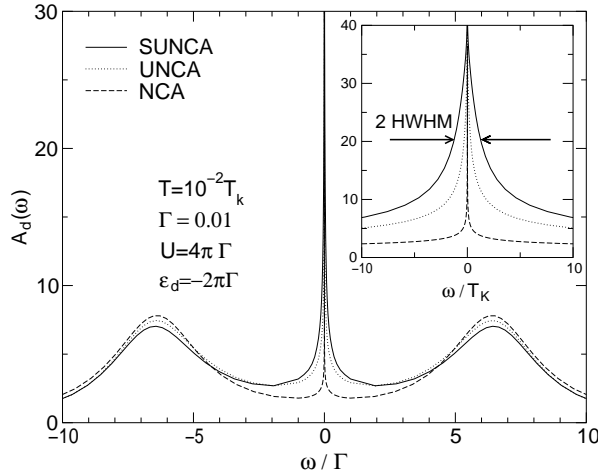,width=0.44\linewidth}

\vspace*{-0.345\linewidth}
\hspace*{0.48\linewidth}\psfig{figure=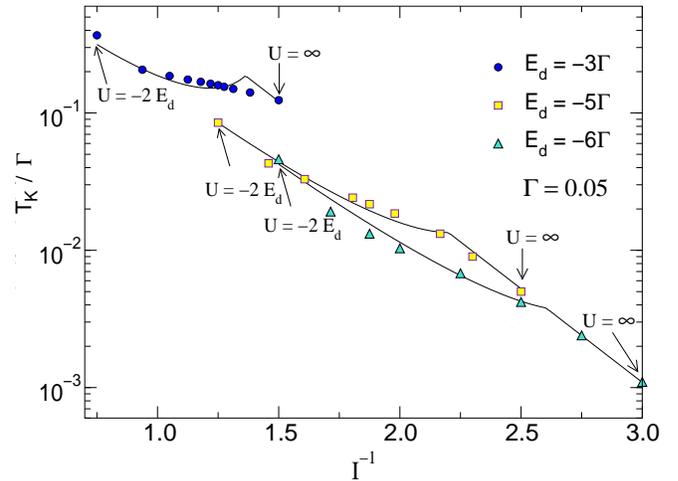,width=0.48\linewidth}
\vspace*{0.6cm}
\caption{Left panel: 
Local electron spectral function calculated using
NCA, UNCA, and SUNCA. The Kondo temperature is determined by the
peak width. It is seen that in NCA it comes out orders of magnitude
too low. --- Right panel: Kondo Temperature for various parameters
$E_d$, $U$ and fixed $\Gamma$. Solid lines represent the results from
the Bethe ansatz solutions, Eq.\ (99). Data points are the SUNCA results
determined from the width of the Kondo peak in the d-electron
spectral function.   
\label{AdU}}
\end{figure}

The self--consistent set of equations defined 
by Fig.~\ref{Uselfenergy} and the
expression for the pseudo-particle Green's function, Eq.~(\ref{green}), 
supplemented by the expression for the heavy boson Green's function
\begin{equation}
{\cal G}_{a}(i\omega_n) = \Big\{ i\omega_n - (2E_d +U) -
\Sigma_{a}(i\omega_n)\Big\}^{-1} \ ,
\label{greena}
\end{equation}
have been solved numerically \cite{haule00,haule01}. As
expected, the pseudoparticle
spectral functions are found to display power law divergencies at
the infrared threshold. 
The exact power law exponents, 
derived from the Friedel sum rule argument based on the assumed Fermi
liquid ground state (section III B), are given by 
\begin{eqnarray}
\alpha_f &=& n_d - \frac{n_d^2}{2}  \\
\alpha_b &=& 1-\frac{n_d^2}{2} \\
\alpha_a &=& -1 +2n_d -\frac{n_d^2}{2} 
\label{eq:exponents}
\end{eqnarray}
The threshold exponents obtained from the SUNCA solution 
are shown in Fig.~\ref{exponentsU} as functions of the average impurity
occupation number $n_d$. They agree with the
exact results in the Kondo limit ($n_d \to 1$) where $\alpha_f =
\alpha_b = \alpha_a = \frac{1}{2}$, but deviate
for $n_d < 1$.
It is expected that the SUCTMA will recover the
correct exponents, as does the CTMA in the case of infinite $U$.
The d-electron spectral function for the symmetric Anderson model as
obtained in SUNCA is shown in Fig.~\ref{AdU} (left panel). 
In this case the Kondo
resonance is located exactly at $\omega = 0$.  Its width is a measure
of the Kondo temperature $T_K$.  The three curves shown correspond to
the elementary NCA (Fig.~\ref{SUNCA} a)), the UNCA (Fig.~\ref{SUNCA} b)) 
and the SUNCA.  The
inset shows that for the parameters chosen the $T_K$ obtained in NCA
is too small by a 
\noindent
factor $10^{-2}$, whereas the UNCA is only a factor
of $1/3$ off.  Contrary to Ref.~\cite{pruschke89} we
do not find that the UNCA is sufficient to recover $T_K$.  
In Fig.~\ref{AdU} (right panel) 
the results of the SUNCA for $T_K$ as determined from the
width of the Kondo resonance are compared to the exact result, 
\begin{equation}
T_K = {\rm min}\Big\{ \frac{1}{2\pi}U \sqrt{I},\; \sqrt{D\Gamma}
\Big\} {\rm e}^{(-\pi/I)} \ ,
\label{7.4}
\end{equation}
where $I = \frac{2\Gamma U}{\mid E_d\mid(U+E_d)}$.  The agreement is
seen to be very good for a large range of parameters $I$ and $E_d$.

\section{Conclusion}
We have reviewed a technique
to describe correlated quantum impurity systems with strong onsite
repulsion, which is based on a conserving formulation of the 
auxiliary boson method. 
The conserving scheme allows to implement the
conservation of the local charge $Q$ without taking into account 
time dependent fluctuations of the gauge field $\lambda$, while exactly
projecting the quantum dynamics onto the physical subspace of no
double occupancy of sites (in the limit of infinitely strong 
on-site repulsion). Any spurious condensation of the auxiliary boson field 
is avoided in this way. 

By including the physically dominant contributions, 
spin flip processes in the Kondo regime, and 
spin as well as charge fluctuation processes in the mixed valence
and empty impurity regimes, the method recovers the Fermi liquid
ground state of the single-channel Anderson impurity model as well as the
non-Fermi-liquid low-temperature behavior of the two-channel
Anderson model: The correct infrared threshold exponents of the auxiliary
particle propagators, which are identified as indicators for Fermi or
non-Fermi liquid behavior, are obtained in both cases.
Physical quantities, like the magnetic susceptibility, 
are correctly described both in the Fermi and in the non--Fermi liquid 
cases of the model
over the complete temperature range, including the crossover to the 
correlated many--body state at the lowest temperatures, which 
has previously been notoriously difficult to obtain within a diagrammatic
many--body theory.

\widetext
\narrowtext

We have also presented a generalization of the auxiliary particle 
method for finite on-site repulsion $U$ and solved the corresponding
equations on the level of a generalized non-crossing approximation (SUNCA),
where virtual excitations into the empty or into the doubly occupied
impurity state are treated in a fully symmetric way. In this way the
correct low-temperature scale of the finite-U Anderson model was
obtained. The controlled treatment of 
a finite on-site repulsion 
is a precondition for applying the auxiliary particle method to models of 
the Hubbard type or -- within the scheme of the dynamical mean field
theory (limit of infinite dimensions) -- of the Anderson impurity type, 
where a Mott--Hubbard metal--insulator
transition occurs as a function of band filling for finite $U$.
As a standard diagram technique the conserving auxiliary particle method
has the potential to be 
applicable to problems of correlated systems on a lattice as well as
to mesoscopic systems out of equilibrium via the Keldysh technique.
These developments are currently under investigation.

Collaboration and discussions with S. B\"ocker, T. A. Costi,
K. Haule, S. Kirchner, and Th. Schauerte are gratefully acknowledged.
This work is supported in part by DFG through SFB 195 and by
a grant of computer time of the High-Performance Computing 
Center Stuttgart.

\end{multicols}

\vspace*{1cm}

\begin{appendix}
\narrowtext
\section*{Infrared cancellation of non--CTMA diagrams} 
The CTMA is not only justified on physical grounds by the inclusion 
of the maximum number of spin flip and charge fluctuation processes 
at any given order of perturbation theory, but also by an infrared
cancellation of all diagrams not included in the CTMA. In the following
we will prove this cancellation theorem.

\begin{figure}
\vspace*{-0cm}
\centerline{\psfig{figure=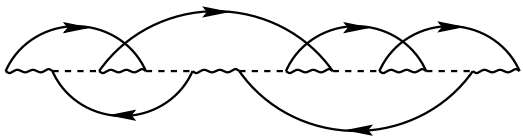,width=0.8\linewidth}}
\vspace*{0.6cm}
\caption{
Typical pseudofermion self--energy skeleton diagram of loop order
$L=6$, containing $L_{sp}=1$ spin (or fermion) loop and $L_{ch}=1$
channel loop. 
\label{selfenergy}}
\end{figure}

(1) {\em Power counting}.---
Each auxiliary particle loop carries a factor of the fugacity 
${\rm exp}(-\beta\lambda)$, which vanishes upon projection onto the
$Q=1$ subspace, $\lambda\rightarrow \infty$. Therefore, an arbitrary 
$f$ or $b$ self--energy diagram consists of one single line of 
alternating fermion and boson propagators, with the hybridization 
vertices connected by conduction electron lines in any possible way, 
as shown in Fig.~\ref{selfenergy} (see also \cite{keiter.71}). 
Such a fermion self--energy skeleton diagram of loop order $L$ is 
calculated as
\widetext
\top{-2.8cm}
\begin{eqnarray}
\Sigma _{f}^{(L)}(\omega ) &=& (-
1)^{3L-1+L_{sp}}\;N^{L_{sp}}\;M^{L_{ch}}\;
\Gamma ^L
\int\frac{ d\varepsilon _1}{\pi}\dots \frac{ d\varepsilon _L}{\pi}
\;f(\varepsilon _1) \dots \;f(\varepsilon _L) \
A^0_{c}(s_1\varepsilon _1)\dots A^0_{c}(s_L\varepsilon _L)
\label{eq:selfenergy}\\
&\times&\;G_b(\omega+\omega _1)G_f(\omega+\omega_1+\omega_1')\;
\dots\;G_b\Big(\omega+\sum _{i=1}^{k}\omega _i+
                      \sum _{i=1}^{k-1}\omega _i'\Big)\;
       G_f\Big(\omega+\sum _{i=1}^{k}(\omega_i+\omega_i')\Big)\;\dots\;
       G_b(\omega+\omega _L)\; ,\nonumber
\end{eqnarray}
\bottom{-2.7cm}
\narrowtext
where $G_{f,b}$ are the {\em renormalized}\/, i.e. power law divergent 
auxiliary particle propagators, and
$L_{sp}$ and $L_{ch}$ denote the number of spin (or fermion, $c$--$f$) 
loops and the number of channel (or $c$--$b$) loops contained in the
diagram, respectively. Spin and channel indices are not shown
for simplicity. Each of the auxiliary particle 
frequencies $\omega _i$, $\omega _i'$ coincides
with one of the integration variables $\varepsilon _j$, $j=1,\dots , L$, 
in such a way that energy is conserved at each hybridization vertex.
This implies that the sign of the frequency carried by
a $c$--electron line is $s_i = +$, if the $c$--electron line runs from
right to left, and $s_i = - $, if it runs from left to right in 
Fig.~\ref{selfenergy}. An analogous expression holds for the
slave boson self--energy diagrams.
\noindent
By substituting 
$x_j = \varepsilon _j/\omega$, $j=1,\dots , L$ and factoring 
out $\omega ^{-\alpha_{f,(b)}}$ from each fermion (boson) propagator, 
the infrared behavior of the term Eq.~(\ref{eq:selfenergy}) 
is deduced as
\begin{eqnarray}
{\rm Im}
\Sigma _{f,b}^{(L)}(\omega ) = 
C \omega ^{\alpha _{f,b}+L(1-\alpha _f-\alpha _b)},
\label{eq:powercount}
\end{eqnarray}
where $C$ is a finite constant. Clearly, when the NCA solutions 
are inserted for the propagators $G_{f,b}$, 
i.e. $\alpha _f +\alpha _b = 1$, their power law behavior is just 
reproduced by any term of the form Eq.~(\ref{eq:selfenergy}). 
However, this is no longer the case for the exact propagators 
in the Fermi liquid regime ($M < N$), where in 
general $\alpha _f +\alpha _b > 1$.  
Thus, the infinite resummation of terms to arbitrary loop order is
unavoidable in this case.

(2) {\em Infrared cancellation}.--- As discussed in section VI B, the CTMA
is equivalent to the self--consistent summation of all skeleton free
energy diagrams, where a conduction electron line spans at most two
hybridization vertices (Fig.~\ref{CTMA}). 
Thus, any skeleton self--energy diagram {\em not}\/
included in CTMA contains at least one conduction electron ``arch''
which spans four (or more) vertices, with 
\begin{figure}
\vspace*{-0cm}
\centerline{\psfig{figure=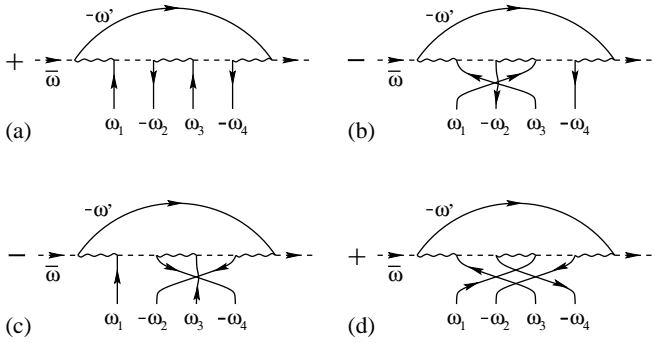,width=1.0\linewidth}}
\vspace*{0.7cm}
\caption{
Set of contributions to skeleton diagrams {\em not}\/
contained in CTMA which cancel in the infrared limit
to leading and subleading order in the external frequency, 
$\omega \rightarrow 0$.   
\label{cancellation}}
\end{figure}
\noindent
four conduction lines reaching
from inside to outside of the arch as shown in Fig.~\ref{cancellation}(a).
For each such diagram there exists another skeleton, which differs 
from Fig.~\ref{cancellation}(a) only in that the end points of two 
conduction lines inside the 
arch are interchanged 
(Fig.~\ref{cancellation}(b)). The corresponding permutation of
fermionic operators implies a relative sign between the terms
Fig.~\ref{cancellation}(a) and (b). Without loss of 
generality we now assume
$\omega >0$ for the external frequency of the self--energy.
The leading infrared singular behavior of the term (\ref{eq:selfenergy})
arises from those parts of the integrations, where the arguments of
the $G_f$, $G_b$ are such that the divergences of all propagators lie
within the integration range. This implies at least 
$-\omega \leq \varepsilon _j \leq 0$, $j=1,\dots , L$. Therefore,
the terms corresponding to Fig.~\ref{cancellation}(a), (b)
differ only in the frequency arguments of the Green's functions inside
the arch, and at temperature $T=0$ the leading infrared behavior of their 
sum reads,
\widetext
\top{-2.8cm}
\begin{eqnarray}
\Sigma _{f}^{(L,a)}(\omega )+\Sigma _{f}^{(L,b)}(\omega ) 
&{\buildrel{_{\omega \to 0}}\over{=}}& 
(-1)^{3L-1+L_{sp}}\;N^{L_{sp}}M^{L_{ch}}
\Gamma ^L \times \label{eq:cancel} \\
&&\hspace*{-3.8cm}
\int _{-\omega}^0 \frac{ d\varepsilon _1}{\pi}
\dots\frac{ d\varepsilon _L}{\pi}\;F(\omega,\{\varepsilon _j\})\;
\big[ G_f(\bar\omega+\omega '+\omega _1)
      G_b(\bar\omega+\omega '+\omega_1+\omega_2)
-G_f(\bar\omega+\omega '+\omega _3)
 G_b(\bar\omega+\omega '+\omega_3+\omega_2)
\big]\nonumber
\end{eqnarray}
\bottom{-2.7cm}
\narrowtext
Here $\bar\omega$ denotes the sum of all frequencies $\omega$,
$\varepsilon _j$ entering the diagrammatic part, 
Fig.~\ref{cancellation}, from the left, and $F(\omega,\{\varepsilon _j\})$ 
consists of all terms which are not altered by interchanging the
$c$--electron lines. In the infrared limit, $\omega _1-\omega _3 \to 0$,
the term in square brackets may be written as
\begin{eqnarray}
\frac{d}{d\bar\omega}\big[G_f(\bar\omega+\omega ')
                          G_b(\bar\omega+\omega '+\omega_2)\big]
                     (\omega _1-\omega _3)
\end{eqnarray}
and upon performing the integrations 
over $\omega _1$, $\omega _3$ the difference $(\omega _1-\omega _3)$ leads
to an additional factor of $\omega $.  
A similar cancellation of the leading infrared singularity occurs
between the terms shown in Fig.~\ref{cancellation}(c), (d).
In an analogous way it may be shown that combining the 
terms Fig.~\ref{cancellation} (a)--(d) leads to a factor of $\omega^2$
compared to the power counting result for one single term.
Thus, the infrared singularity of all non--CTMA terms of loop order $L$
is weaker than the $L$th order CTMA terms by at least $O(\omega ^2)$,
\begin{equation}
\Sigma _{f,b}^{(L,a)}(\omega )+\dots +\Sigma _{f}^{(L,d)}(\omega ) 
\ {\buildrel{_{\omega \to 0}}\over{\propto}}\ 
\omega ^{\alpha _{f,b}+L(1-\alpha _f-\alpha _b)+2}.
\end{equation}
It should be emphasized that in the above derivation, $L$ appears only
as a parameter and, thus, the cancellation theorem holds for 
arbitrarily high loop order $L$. This proves that the CTMA captures the 
leading and subleading infrared singularities ($\omega \to 0$)
at any given order $L$.
\end{multicols}
\end{appendix}

\vspace*{1cm}

\narrowtext

\end{multicols}

\begin{thebibliography}{99} 
\bibitem{Lee86} P.A. Lee, T.M. Rice,
J.W. Serene, L.J. Sham, and J.W. Wilkins, Comments on Condensed Matter
Physics {\bf 12,} 99 (1986).  
\bibitem{Ander97} P.W. Anderson, Science
{\bf 235}, 1196 (1987); and ``The Theory of Superconductivity in the
High-T$_c$ Cuprates'', (Princeton University Press, 1997).
\bibitem{Sarma96} For a review see: ``Perspectives in Quantum Hall
Effects'', eds. S. Das Sarma and A. Pinczuk (Wiley, New York, 1996).
\bibitem{Abrik65} A.A. Abrikosov, Physics {\bf 2}, 21
(1965).  
\bibitem{Barnes76} S.E. Barnes, J. Phys. F {\bf 6}, 1375
(1976); {\bf 7}, 2637 (1977).  
\bibitem{Coleman84} P. Coleman,
Phys. Rev. B {\bf 29}, 2 (1984).  
\bibitem{Newns83} D.M. Newns,
N. Read, J. Phys. C {\bf 16}, 3273 (1983); Adv. Phys. {\bf 36}, 799
(1988).
\bibitem{Hewson93} A.C. Hewson, ``The Kondo Problem to Heavy
Fermions'' (Cambridge University Press, 1993).  
\bibitem{Cox98}
D.L. Cox and A. Zawadowski, Adv. Phys. {\bf 47}, 599 (1998).
\bibitem{schoen97} See, e.g., related articles in
{\it Mesoscopic Electron Transport}, L.~L.~Sohn, L.~P.~Kouwenhoven
and G.~Sch\"on eds., NATO ASI Series E: Applied Sciences {\bf 345}
(Kluwer Science Publishers, Dordrecht, Boston, London, 1997).  
\bibitem{Ander61} P.W. Anderson, Phys. Rev. {\bf 124}, 41 (1961).
\bibitem{Schrief66} J.R. Schrieffer and P.A. Wolff, Phys. Rev. {\bf
149}, 491 (1966).  
\bibitem{Wilson75} K.G. Wilson,
Rev. Mod. Phys. {\bf 47}, 773 (1975).  
\bibitem{Andrei83} N. Andrei,
K. Furuya and J.H. L\"owenstein, Rev. Mod. Phys. {\bf 55}, 331 (1983).
\bibitem{Wieg83} P.B. Wiegmann and A.M. Tsvelik, J. Phys. C {\bf 12},
2281 and 2321 (1983).  
\bibitem{Nozieres74} P. Nozi\'eres, J. Low
Temp. Phys. {\bf 17}, 31 (1974).  
\bibitem{Andrei84} N. Andrei and
C. Destri, Phys. Rev. Lett. {\bf 52}, 364 (1984).  
\bibitem{Tsvelik85}
A.M. Tsvelik, J. Phys.C {\bf 18}, 159 (1985).  
\bibitem{affleck.91}
I.~Affleck and A.W.W.~Ludwig, Nucl.~Phys.~{\bf 352}, 849, (1991); {\bf
B360}, 641, (1991); Phys.~Rev.~B {\bf 48}, 7297 (1993).
\bibitem{yamada.75} K.~Yamada, Prog. Theor. Phys. {\bf 53}, 970
(1975); {\em ibid.}\/ {\bf 54}, 316 (1975); {\em ibid.}\/ {\bf 55},
1345 (1976); K.~Yosida and K.~Yamada, Prog. Theor. Phys. {\bf 53},
1286 (1975).
\bibitem{coxwilkins.87} D. L. Cox, N.~E.~Bickers and J. W. Wilkins,
Phys. Rev. B {\bf 36}, 2036 (1987).
\bibitem{note1}
This means that the auxiliary particle propagators 
are {\em not}\/ calculated in the canonical ($Q=1$) ensemble. 
The projection onto the $Q=1$ sector of Fock space is achieved only 
when they are combined to calculate expectation values of physically 
observable operators like $G_{d\sigma}$, $\langle\vec S\rangle$ etc.
The latter can be seen explicitly, e.g., from Eq.~(\ref{gdNCA}), 
2nd equality.
\bibitem{anderson.67} P. W. Anderson, Phys. Rev. Lett.  {\bf 18}, 1049
(1967).
\bibitem{nozieres.69} P. Nozi\`eres and C. T. De Dominicis,
Phys. Rev. {\bf 178}, 1073; 1084; 1097 (1969).
\bibitem{mahan.80} G. D. Mahan, {\em Many--Particle Physics}, 2nd ed.,
pp. 732 (Plenum Press, New York, 1990) gives an overview.
\bibitem{schotte.69} K. D. Schotte and U. Schotte, Phys. Rev. {\bf
185}, 509 (1969).
\bibitem{mengemuha.88} B.~Menge and E.~M\"uller--Hartmann,
Z.~Phys.~{\bf B73}, 225 (1988).

\bibitem{costi.94} T.A. Costi, P. Schmitteckert, J. Kroha and
P. W\"olfle, Phys. Rev. Lett. {\bf 73}, 1275 (1994).

\bibitem{costi.94b} T.A. Costi, P. Schmitteckert, J. Kroha and
P. W\"olfle, Physica (Amsterdam) {\bf 235--240C}, 2287 (1994).

\bibitem{fujimoto.96} S. Fujimoto, N. Kawakami and S.K. Yang,
J.Phys.Korea {\bf 29}, S136 (1996).
\bibitem{readnewns.88} D.~M.~Newns and N.~Read,
                      J.~Phys.~{\bf C16}, 3273 (1983); 
                      Adv.~Phys.~{\bf 36}, 799 (1988).

\bibitem{elitzur.75} S. Elitzur, Phys. Rev. D {\bf 12},3978 (1975). 

\bibitem{jevicki.77} A. Jevicki, Phys. Lett. B {\bf 71}, 327 (1977).

\bibitem{david.81} F. David, Comm. Math. Phys. {\bf 81}, 149 (1981).

\bibitem{lawrie.85} I.D. Lawrie, J. Phys. A {\bf 18}, 1141 (1985).

\bibitem{kuroda.88} B.~Jin and Y.~Kuroda, J.~Phys.~Soc.~Japan {\bf 57},
                    1687 (1988).

\bibitem{kuroda.97} T.~Matsuura et al.,
                    J.~Phys.~Soc.~Japan {\bf 66}, 1245 (1997). 

\bibitem{sellier98} G. Sellier, Diploma Thesis, Universit\"at Karlsruhe,
unpublished (1998).

\bibitem{kadanoff.61} G.~Baym and L.P.~Kadanoff, Phys.~Rev. {\bf 124},
                 287 (1961); G.~Baym, Phys.~Rev. {\bf 127} 1391 (1962).

\bibitem{keiter.71} H. Keiter and J. C. Kimball, J. Appl. Phys. {\bf 42},
1460 (1971); N. Grewe and H. Keiter, Phys. Rev B {\bf 24}, 4420 (1981).

\bibitem{kuramoto.83} Y. Kuramoto, Z. Phys. B {\bf 53}, 37 (1983);
         H. Kojima, Y. Kuramoto and M Tachiki,  
                      {\em ibid.}\/ {\bf 54}, 293 (1984);
         Y. Kuramoto and H. Kojima, {\em ibid.}\/ {\bf 57}, 95, (1984);
         Y. Kuramoto, {\em ibid.}\/ {\bf 65}, 29, (1986).


\bibitem{muha.84} 
       E. M\"uller--Hartmann, {\em Z. Phys.} {\bf B57}, 281 (1984).

\bibitem{costi.96} 
T.A.~Costi, J.~Kroha, and P.~W\"olfle, PRB~{\bf 53}, 1850 (1996).


\bibitem{bickers.87} N. E. Bickers, {\em Rev. Mod. Phys.} {\bf 59},
              845 (1987); N. E. Bickers, D. L. Cox \& J. W. Wilkins, 
                  {\em  Phys. Rev.}  {\bf  B36}, 2036 (1987).

\bibitem{coxruck.93} D.~L.~Cox and A.~E.~Ruckenstein, 
       Phys.~Rev.~Lett.~{\bf 71}, 1613 (1993).

\bibitem{anders.94} F.~Anders and N.~Grewe, Europhys.~Lett.
{\bf 26}, 551 (1994); F.~Anders, J.~Phys.~Cond.~Mat.~{\bf 7}, 2801 (1995).

\bibitem{kroha.97} J.~Kroha, P.~W\"olfle and T.~A.~Costi,
                   Phys.~Rev.~Lett.~{\bf 79}, 261 (1997). 

\bibitem{schauerte.00}
Th.~Schauerte, J.~Kroha, and P.~W\"olfle,
Phys.~Rev.~B {\bf 62}, 4394 (2000). 

\bibitem{kirchner.01}
S.~Kirchner, J.~Kroha, and P.~W\"olfle, in preparation.

\bibitem{pruschke89}
Th. Pruschke and N.~Grewe, Z.~Phys.~B {\bf 74}, 439 (1989).

\bibitem{holmschoenh89}
I.~Holm and K.~Sch\"onhammer, Solid State Comm. {\bf 69}, 10,
969 (1989).


\bibitem{haule00}
K.~Haule, S. Kirchner, J.~Kroha, and P.~W\"olfle, 
in {\it Proceedings of the NATO Advanced Research Workshop on
``Size Dependent Magnetic Scattering''}, Pecs, Hungary 2000
(Kluwer Academic Publishers, in press).

\bibitem{haule01}
K.~Haule, S. Kirchner, J.~Kroha, and P.~W\"olfle, cond-mat/0105490 (2001).
\end{thebibliography}
\end{document}